\documentclass[%
aps,
reprint,
superscriptaddress,
noeprint,
amsmath,amssymb,
prl, %
floatfix,nofootinbib
]{revtex4-2}

\usepackage{chemformula}
\usepackage[utf8]{inputenc}
\usepackage[T1]{fontenc}
\usepackage{graphicx}
\usepackage{siunitx}
\usepackage{amsmath}
\usepackage{amssymb}
\usepackage[colorlinks=true,allcolors=blue]{hyperref}
\usepackage{soul}
\usepackage[normalem]{ulem}
\usepackage{lipsum}
\usepackage{xcolor}
\usepackage{pdfpages,pgffor}

\newcommand{\orcid}[1]{\href{https://orcid.org/#1}{\includegraphics[width=8pt]{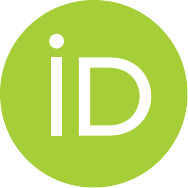}}}
\makeatletter
\AtBeginDocument{\let\LS@rot\@undefined}
\makeatother

\begin{document}
\title{One-Dimensional Moiré Physics and Chemistry in Heterostrained Bilayer Graphene} 

\author{Gabriel R. Schleder\orcid{0000-0003-3129-8682}}
\email[E-mail: ]{gabriel.schleder@lnnano.cnpem.br}
\thanks{These authors contributed equally}
\affiliation{John A. Paulson School of Engineering and Applied Sciences, Harvard University, Cambridge, MA 02138, United States}
\affiliation{Brazilian Nanotechnology National Laboratory (LNNano), CNPEM, 13083-970 Campinas, São Paulo, Brazil}

\author{Michele Pizzochero}
\email[E-mail: ]{mpizzochero@g.harvard.edu}
\thanks{These authors contributed equally}
\affiliation{John A. Paulson School of Engineering and Applied Sciences, Harvard University, Cambridge, MA 02138, United States}

\author{Efthimios Kaxiras}
\email[E-mail: ]{kaxiras@g.harvard.edu}
\affiliation{John A. Paulson School of Engineering and Applied Sciences, Harvard University, Cambridge, MA 02138, United States}
\affiliation{Department of Physics, Harvard University, Cambridge, MA 02138, United States}

\begin{abstract}
Twisted bilayer graphene (tBLG) has emerged as a promising platform to explore exotic electronic phases. However, the formation of moiré patterns in tBLG has thus far been confined to the introduction of twist angles between the layers. Here, we propose heterostrained bilayer graphene (hBLG), as an alternative avenue to access twist-angle-free moiré physics via lattice mismatch. Using atomistic and first-principles calculations,  we demonstrate that uniaxial heterostrain can promote isolated flat electronic bands around the Fermi level. Furthermore, the heterostrain-induced out-of-plane lattice relaxation may lead to a spatially modulated reactivity of the surface layer, paving the way for the moiré-driven chemistry and magnetism. We anticipate that our findings can be readily generalized to other layered materials.
\end{abstract}

\maketitle

\begin{figure*}[!ht]
\centering
\includegraphics[width=\linewidth]{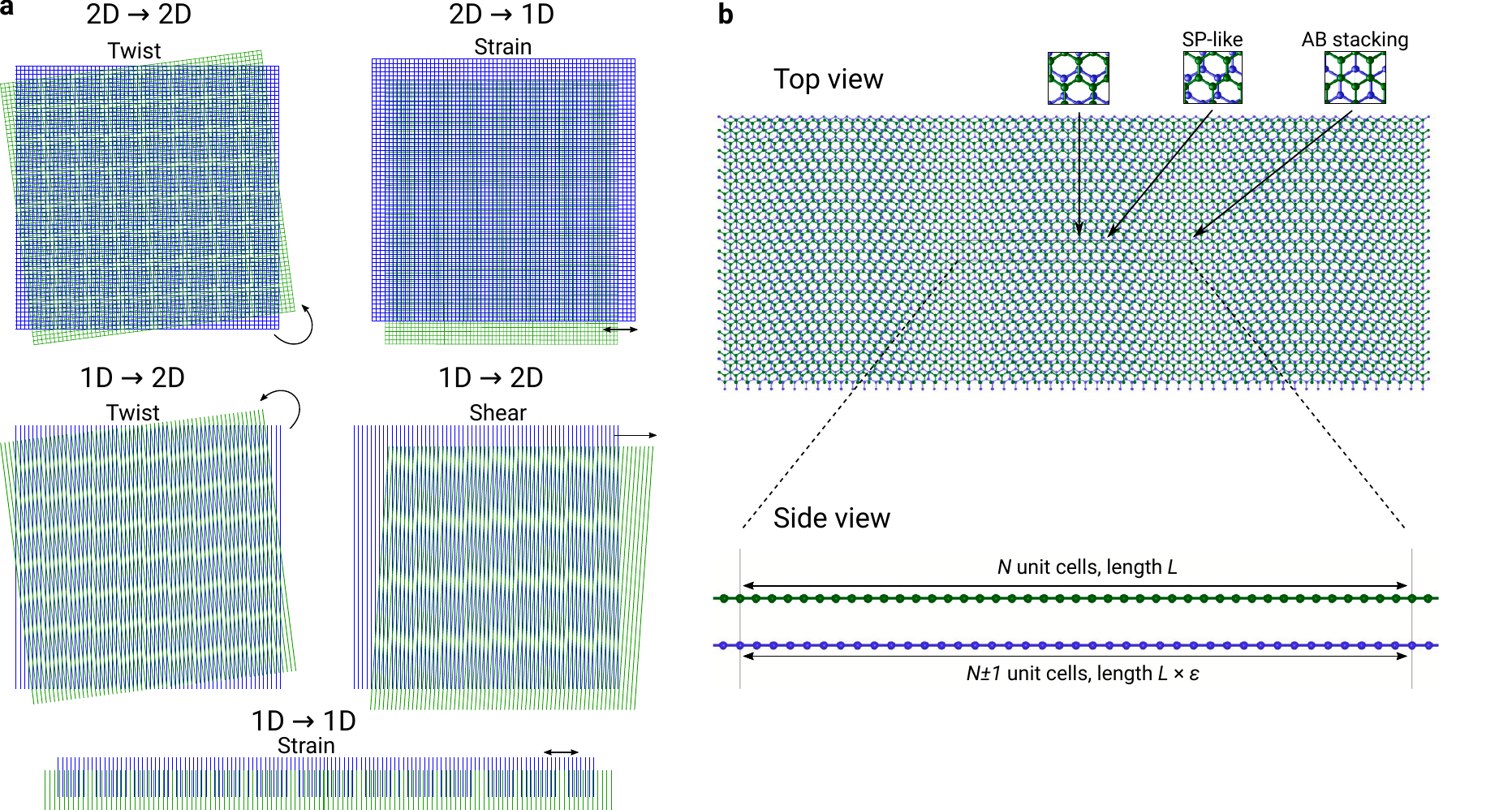}
\caption{Different types of moiré patterns. \textbf{a)} Examples of mechanisms that lead to moiré patterns in one and two dimensions.
\textbf{b)} The supercell matching approach used in this work to model heterostrained bilayers.
The local stackings are highlighted in the insets, periodically changing from AB to displaced AA-like regions, separated by domain walls or saddle-point-like (SP-like) regions.}
\label{fig:moire_types}
\end{figure*}

\paragraph{Introduction.}
The stacking of a pair of graphene layers with a relative twist angle between them introduces an in-plane moiré pattern and a consequent modulation of the lattice potential. Depending on the magnitude of the twist angle, the resulting twisted bilayer graphene (tBLG) exhibits a wide range of intriguing electronic properties, such as superconductivity, correlated insulating and topological semimetallic phases \cite{Cao2018_sc,Cao2018_corr,Dean2013}. These observations have sparked considerable interest in the investigation of the physical and chemical properties of moiré materials, an emerging field dubbed ``twistronics'' \cite{Carr2017_twistronics,Tritsaris2021,Andrei2021,Angeli2022,Kennes2021}. 
In addition to twisting, there are several means to create moiré patterns in 2D materials, such as shear and strain. Furthermore, it is possible to achieve one-dimensional moiré patterns between 2D materials by enforcing a periodic modulation selectively in one direction, as illustrated in Fig. \ref{fig:moire_types}a.
Different approaches to achieve these moiré patterns have been proposed, including vertical and lateral heterostructures of lattice-mismatch heterostrained materials \cite{Waters2020,Salvato2022,Rakib2022,Li2023}, nanotubes \cite{Bonnet2016,ArroyoGascn2020}, substrate-driven moiré patterns, or chemically functionalized surfaces \cite{Li2021,Gupta2022,Yang2023}.
However, earlier theoretical \cite{Cosma2014,Bi2019,Zhu2018,Kogl2023} and experimental \cite{Bai2020,Gao2021,Mesple2021,Zhang2022,Kapfer2023} works restrict their focus mainly on the twist angle as a knob to tune moiré materials, with the individual effect and properties of heterostrain currently lacking a detailed understanding.

In this Letter, we expand the class of moiré materials beyond twisted systems by introducing heterostrained bilayer graphene (hBLG), where uniaxial strain is exerted only in one of the two layers. This heterostrain creates a 1D moiré pattern in the absence of a twist angle. Using first-principles and atomistic simulations, we demonstrate that hBLG can feature isolated electronic bands with vanishing dispersion near the Fermi level, similarly to twisted bilayer graphene near the magic angle. Additionally, we show that the out-of-plane lattice relaxation induced by heterostrain results in a moiré chemical reactivity of the surface layer. 
Our findings establish heterostrained bilayer graphene as a platform to access and engineer moiré physics in 2D materials without relying on twist angles.

\begin{figure}[!h]
\centering
\includegraphics[width=1\columnwidth]{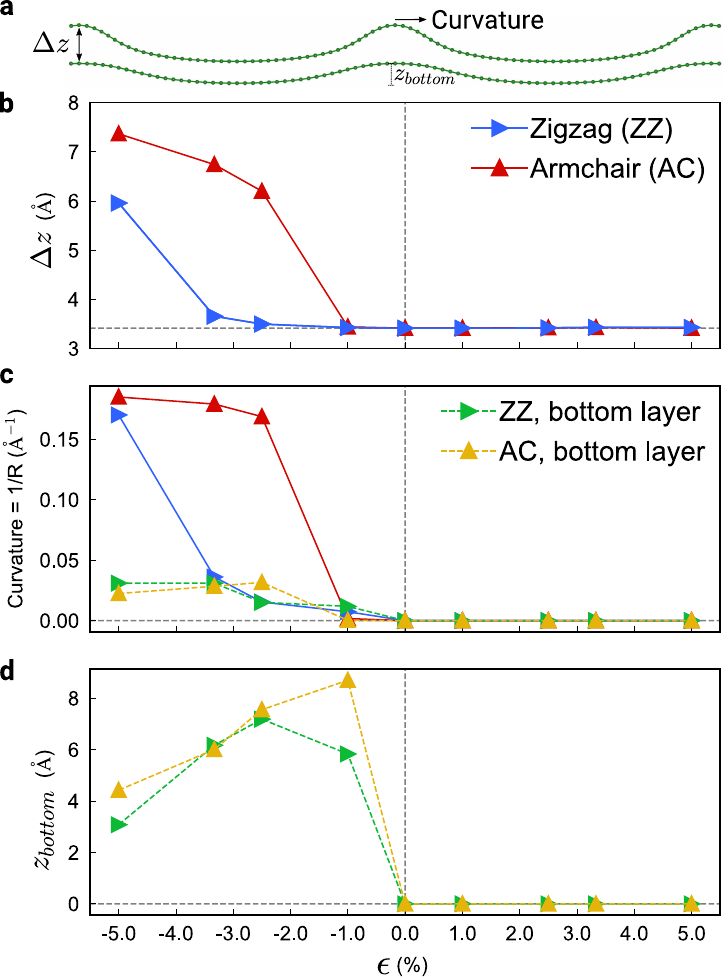}\\
\caption{Structural properties of hBLG. \textbf{a)} Side-view of heterostrained bilayer graphene for a representative heterostrain  value of $\epsilon =$ -5 \%; also shown are the lattice sites of maximum curvature and interlayer distance. \textbf{b)} Maximum interlayer distance $\Delta z$ and \textbf{c)} layer curvature as a function of the heterostrain applied. Positive and negative values correspond to tensile and compressive strains, respectively. Structural properties of the unstrained bottom layer: \textbf{d)} the $z$-axis difference, corresponding to the layer height, and the \textbf{e)} layer curvature.}
\label{fig:structural_properties}
\end{figure}

\paragraph{Structural relaxation in hBLG.} In Fig.\ \ref{fig:moire_types}b, we illustrate the supercell-matching approach that we adopt to generate heterostrained models. Although we focus on graphene, this approach can be used for any periodic crystal. We start with an unstrained rectangular cell of graphene and repeat it $N$ times in the direction along which strain will be introduced. For the strained layer, we repeat it $N \pm 1$ times, such that this layer is strained to match the lattice constant of the unstrained one. The resulting model will thus have a strain of $\varepsilon=\pm 1/N$.
The strain can be exerted either in the zigzag or armchair direction. 
We optimize the atomic structures by means of a registry-dependent atomistic force-field potential, as discussed in more detail in Supplementary Note 1.
From the optimized atomic structure, we determine the maximum interlayer distance and external curvature of the unstrained layer. These results are summarized in Fig.\ \ref{fig:structural_properties}, and figures of structures before and after relaxation are presented in the Supporting Note 4.  

The structural relaxation is the result of a competition between the minimization of the local elastic strain and the expansion of the regions exhibiting the lowest-energy interlayer stacking configuration (AB- or Bernal-stacking) and concurrent contraction of regions exhibiting the higher-energy interlayer stacking configuration (AA-stacking). For tensile heterostrain, we observe that the atomic structure retains a planar configuration and features approximately the same interlayer distance as unstrained bilayer graphene. For compressive heterostrain, the atomic structure acquires a pronounced curvature that increases with the magnitude of the heterostrain exerted. For low values of compressive heterostrain, the atomic structure relaxes by homogeneously increasing the curvature, such that the local strain is minimized and therefore their periodic structure has a sinusoidal shape. Above a critical strain value, the homogeneous curvature cannot minimize local forces, resulting in a rippled lattice \cite{Avouris2012,Lin2013,Yazyev2022}, where the strain spatially localizes inducing a local increase of the interlayer distance, as displayed in Fig.\ \ref{fig:structural_properties}a. This regime starts at values of heterostrain of approximately -1\% in the zigzag direction, and at a larger value in the armchair direction, proportional to the ratio of armchair/zigzag lattice vectors length. For both directions, the interlayer distances increase continuously by over 200\%, and the curvature increases from zero (planar) to about 0.17 \AA{}$^{-1}$ (equivalent to a radius of 6 \AA{}).
Interestingly, the unstrained layer accommodates part of the strain, and the transition from the homogenous sinusoidal curvature and the rippled regimes are seen in Fig.\ \ref{fig:structural_properties}c,d, where the curvature increases with increasing compressive strain.

\smallskip

\begin{figure*}[!t]
\centering
\includegraphics[width=1\linewidth]{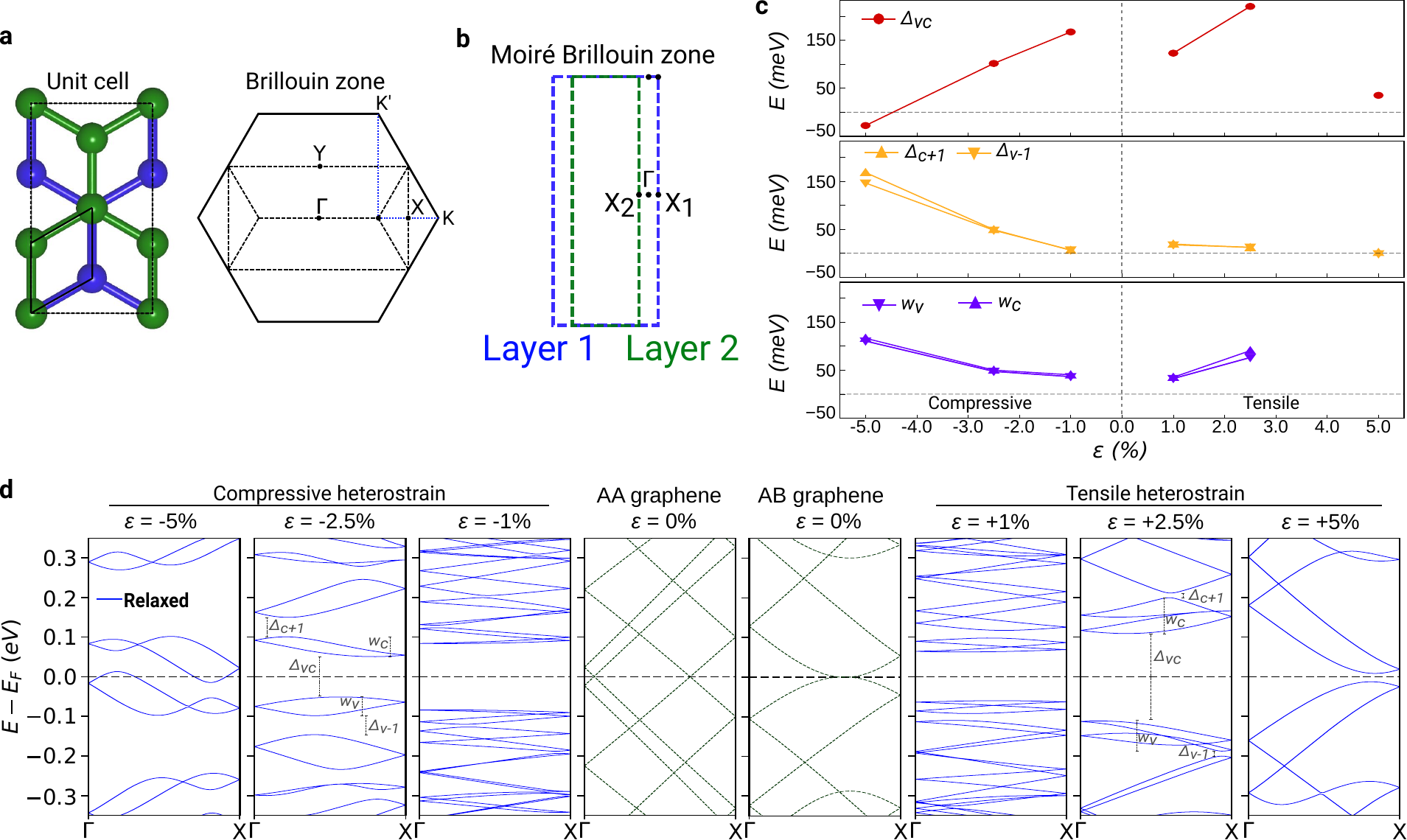}
\caption[font=\columnwidth, font=small]{Electronic structure of hBLG. 
\textbf{a)} Pristine bilayer graphene rectangular cell and Brillouin zone. \textbf{b)} Moiré Brillouin zone obtained from the two individual layers. \textbf{c)} Summary of electronic properties of the relaxed systems: valence--conduction band gap $\Delta_{vc}$, gap to the band below or above it $\Delta_{v-1}$ or $\Delta_{c+1}$, and the width of the valence and conduction moiré bands $w_{v}$ and $w_{c}$.
\textbf{d)} Band structures of hBLGs including relaxation effects. The results for rigid hBLGs, without relaxation effects, are presented in the Supporting Information.
}
\label{fig:electronic_properties}
\end{figure*}

\paragraph{Moiré flat bands in hBLG.} Motivated by the effects of strain engineering in graphene \cite{Pereira2009,Weiss2012,Si2016}, we next investigate the electronic structure of the different hBLG systems. As a reference, in Fig.\ \ref{fig:electronic_properties}a we show the rectangular simulation cell and Brillouin zone of pristine bilayer graphene. In the rectangular cell, the high-symmetry points where the Dirac cones occur, are folded into a single point between the $\Gamma - X$ path \cite{CastroNeto2009,Suzuki2017}. Therefore, to study the low-energy physics of these systems we focus on the effects around this point by means of density functional theory (DFT) calculations, as detailed in Supporting Note 2.
For heterostrain in the \textit{x}-direction (zigzag), the moiré Brillouin zone is displayed in Fig.\ \ref{fig:electronic_properties}b, interpreted as formed by the vectors connecting the high symmetry points of each layer. In Supporting Note 2, we discuss heterostrain applied in the \textit{y}-direction (armchair), which leads to metallic band structures.
Fig. \ref{fig:electronic_properties}c summarizes the main electronic properties for relaxed systems extracted from the band structures: the valence--conduction band gap $\Delta_{vc}$, gap to the band below or above it $\Delta_{v-1}$ or $\Delta_{c+1}$, and the width of the valence and conduction moiré bands $w_{v}$ and $w_{c}$. The results for the rigid systems are presented in the Supporting Information.

In the middle panels of Fig.\ \ref{fig:electronic_properties}d we show the band structures of unstrained AA and AB bilayer graphene in the rectangular supercell ($N=20$) to illustrate band folding effects (the unfolded bands are shown in Supporting Note 2). In line with previous works \cite{CastroNeto2009}, AA-stacked bilayer graphene possesses a pairs of Dirac cones crossing the Fermi energy, while AB-stacked bilayer graphene possesses two sets of parabolic bands, one of which touches at the Fermi level. Farther from the Fermi level, additional crossings are observed due to the folding of these bands.

In Fig.~\ref{fig:electronic_properties}d, we show the band structures of the heterostrained systems including relaxation effects, i.e., the strain is non-uniformly distributed along the strained direction minimizing the elastic energy of the system. The band structures for the rigid systems are presented in the Supporting Information.
The magnitude of the strain ranges from $\varepsilon = -5\%$ (compressive) to $\varepsilon = +5\%$ (tensile).
It has been shown that below a critical strain of 1\%, the heterostrained systems distributes the strain to the second layer in order to maintain a uniform AB-stacking configuration \cite{Georgoulea2022}, while above this critical value the systems have the non-uniform stacking profile seen here.

For both compressive and tensile strain, two sets of isolated valence and conduction bands form, having both a gap between them ($\Delta_{vc}$) and a gap to the band below or above it ($\Delta_{v-1}$ or $\Delta_{c+1}$). 
The strain-induced gaps arise from the potential difference between the layers, resulting in a pseudoscalar effective potential as a transverse electric field \cite{Choi2010}.
The width of these moiré bands ($w_{v}$ and $w_{c}$) decreases with the magnitude of the strain, 
yet features are sensitive to the magnitude and direction of the strain exerted.
For compressive heterostrain, smaller strains lead to flatter bands, and simultaneously $\Delta_{vc}$ increases while $\Delta_{v-1}$ and $\Delta_{c+1}$ decreases, interpreted as valence and conduction bands moving away from each other. Hence, intermediate values of strain lead to isolated bands.
For tensile heterostrain, which retains the planarity of the lattice, larger strains lead to bands similar to AB graphene, while weaker strains also lead to flatter bands, but now simultaneously $\Delta_{vc}$ decreases while $\Delta_{v-1}$ and $\Delta_{c+1}$ increases, interpreted as valence and conduction bands moving closer to each other. Therefore, smaller strains present flatter isolated bands.

In summary, the results for both rigid and relaxed systems shown in Fig.~\ref{fig:electronic_properties}c suggests that both low-tensile and intermediate-compressive heterostrains are promising routes for experimentally achieving flat bands. This is of particular relevance not only to display  spin-charge separation or Peierls distortions, but also to enable strongly correlated physics in one dimension, hosting phases such as Luttinger liquids, band insulators, bond ordered waves, and Mott insulators \cite{Kennes2020}.

\smallskip
\paragraph{Moiré chemistry in hBLG.} 

\begin{figure}[!ht]
\centering
\includegraphics[width=\linewidth]{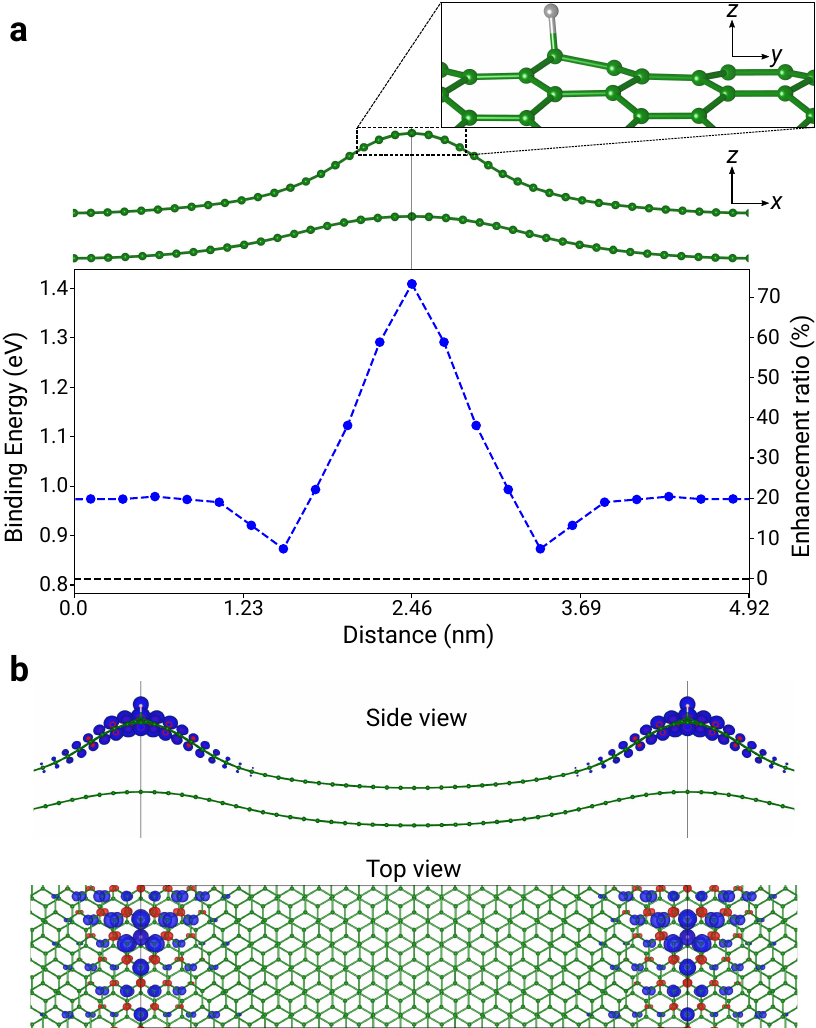}
\caption{Chemical modification of hBLG. \textbf{a)} (Top) Side view illustration of the atomic structure for the hydrogen atom chemisorption on hBLG; the inset shows a perpendicular view with a chemisorbed H atom. (Bottom) Binding energy of an H atom along the moiré pattern of hBLG  compared to monolayer graphene (horizontal dashed line). The horizontal axis matches the atomic structure on top. \textbf{b)} Side and top views of the spin density upon chemisorption of an H atom at the most stable site of hBLG.}
\label{fig:chemistry}
\end{figure}

The heterostrain-induced lattice relaxation in hBLG has important consequences on its surface chemistry. We investigate the  reactivity of hBLG using the first-principles methodology outlined in Supporting Note 3. We consider the adsorption of a hydrogen atom (see Fig.\ \ref{fig:chemistry}a), the simplest yet the most promising ad-species for tailoring the properties of graphene and related nanostructures \cite{Bonfanti2018}. We assess the stability of such adatom at various sites of the surface graphene layer along the one-dimensional moiré pattern. To that end, we define the binding energy as, $E\textsubscript{b}$, as
\begin{equation}
E\textsubscript{b} = (E\textsubscript{Gr} + E\textsubscript{H}) - E\textsubscript{H+Gr},
\end{equation}
where $E\textsubscript{Gr}$ and $E\textsubscript{H+Gr}$ are the total energies of the pristine and hydrogenated nanoribbon, respectively, and $E\textsubscript{H}$ is the energy of an isolated hydrogen atom. According to this expression, positive values of $E\textsubscript{b}$ indicate exothermic processes. 

In Fig.\ \ref{fig:chemistry}a, we give the evolution of $E\textsubscript{b}$ along the moiré pattern of hBLG for the representative model shown in Fig.\ \ref{fig:structural_properties}a, and compare it with the corresponding value for free-standing graphene. The binding energy in hBLG strongly depends on the adsorption site. Its maximum value ($E\textsubscript{b} = 1.41$ eV) occurs at the lattice site of hBLG that exhibits the most pronounced curvature, leading to an enhancement of the stability of the hydrogen adatom by over 75\% compared to free-standing graphene. This is because the local surface curvature induced by the heterostrain weakens the $\pi$-bonds and promotes a facile addition of the hydrogen atom and accompanying $sp^2$ $\rightarrow$ $sp^3$ orbital rehybridization of the functionalized carbon atom \cite{Park2003}. Analogous effects were previously observed for the case of graphene supported on various substrates, such as silicon carbide \cite{Balog2009} or iridium \cite{Balog2010}. Hence, we suggest that the exposure of hBLG to atomic hydrogen may lead to selective hydrogenation of the convex regions, possibly realizing the long-sought patterned chemical functionalization of graphene that has been studied in the context of, e.g., band-gap engineering \cite{Balog2010, Elias2009}.

The addition of the hydrogen adatom introduces a magnetic moment of 1 $\mu\textsubscript{B}$ in otherwise non-magnetic hBLG. Such magnetic moment originates from the removal of a $p_z$ orbital from the $\pi$-electron cloud as a consequence of the formation of the covalent C-H bond. This is in line with Lieb's theorem for the repulsive Hubbard model at half-filling, according to which a lattice imbalance translates to a spin imbalance \cite{Lieb1989}. In Fig.\ \ref{fig:chemistry}b, we show the spin density (i.e., the difference in charge density between spin-majority and spin-minority channels) upon chemisorption of the hydrogen atom at the stable site. Analogously to free-standing graphene, the spin density features a $\sqrt{3} \times \sqrt{3}$ $R$ $30$ localization pattern, as it mainly resides at the majority sublattice and decays away from the adatom \cite{Casolo2009, Yazyev2007, Pizzochero2022}. We thus anticipate that patterned hydrogenation of hBLG may lead to a periodic array of magnetic moments, as illustrated in Fig.\ \ref{fig:chemistry}b, that can be regarded as a spin-half chain hosted in a moiré system \cite{Gonzalez-Herrero2016}.

\smallskip
\paragraph{Summary and Conclusions.} We have proposed heterostrain bilayer graphene (hBLG) as an interesting platform to achieve twist angle-free moiré patterns through the application of uniaxial lattice deformations. 
This could be achieved experimentally by different means, such as using substrates, nanoindentation, mechanical deformations, pressure, piezoelectricity, or having suspended layers \cite{Weiss2012}.
We have shown that fully relaxed hBLG exhibits isolated bands around the Fermi level which become dispersionless at values of strain of approximately 1\%. 
Similar ideas could be exploited for other types of systems, such as metamaterials, photonic crystals \cite{Nguyen2022}, and artificial lattices.
The application of heterostrain leads to a substantial out-of-plane lattice relaxation, which can be exploited to accomplish patterned chemical functionalization \cite{Hsieh2022} of the surface layers through, e.g., hydrogenation, with implications for $\pi$-electron magnetism. To conclude, we anticipate that the approach devised in this work can be qualitatively extended to other van der Waals bilayers, thus opening new avenues for the exploration of moiré physics and chemistry.

\smallskip
\paragraph{Acknowledgments.} 
This research used resources of the National Energy Research Scientific Computing Center (NERSC), operated under Contract No. DE-AC02-05CH11231 using NERSC award BES-ERCAP0020773.
G.R.S. acknowledges funding from the Army Research Office under Cooperative Agreement Number W911NF-21-2-0147.
M.P.\ is financially supported by the Swiss National Science Foundation (SNSF) through the Early Postdoc.Mobility program (Grant No. P2ELP2-191706).
E.K. acknowledges funding from the STC Center for Integrated Quantum Materials, NSF Grant No. DMR-1231319; NSF DMREF Award No. 1922172; and the Army Research Office under Cooperative Agreement Number W911NF-21-2-0147.


\begin{thebibliography}{49}%
\makeatletter
\providecommand \@ifxundefined [1]{%
 \@ifx{#1\undefined}
}%
\providecommand \@ifnum [1]{%
 \ifnum #1\expandafter \@firstoftwo
 \else \expandafter \@secondoftwo
 \fi
}%
\providecommand \@ifx [1]{%
 \ifx #1\expandafter \@firstoftwo
 \else \expandafter \@secondoftwo
 \fi
}%
\providecommand \natexlab [1]{#1}%
\providecommand \enquote  [1]{``#1''}%
\providecommand \bibnamefont  [1]{#1}%
\providecommand \bibfnamefont [1]{#1}%
\providecommand \citenamefont [1]{#1}%
\providecommand \href@noop [0]{\@secondoftwo}%
\providecommand \href [0]{\begingroup \@sanitize@url \@href}%
\providecommand \@href[1]{\@@startlink{#1}\@@href}%
\providecommand \@@href[1]{\endgroup#1\@@endlink}%
\providecommand \@sanitize@url [0]{\catcode `\\12\catcode `\$12\catcode
  `\&12\catcode `\#12\catcode `\^12\catcode `\_12\catcode `\%12\relax}%
\providecommand \@@startlink[1]{}%
\providecommand \@@endlink[0]{}%
\providecommand \url  [0]{\begingroup\@sanitize@url \@url }%
\providecommand \@url [1]{\endgroup\@href {#1}{\urlprefix }}%
\providecommand \urlprefix  [0]{URL }%
\providecommand \Eprint [0]{\href }%
\providecommand \doibase [0]{https://doi.org/}%
\providecommand \selectlanguage [0]{\@gobble}%
\providecommand \bibinfo  [0]{\@secondoftwo}%
\providecommand \bibfield  [0]{\@secondoftwo}%
\providecommand \translation [1]{[#1]}%
\providecommand \BibitemOpen [0]{}%
\providecommand \bibitemStop [0]{}%
\providecommand \bibitemNoStop [0]{.\EOS\space}%
\providecommand \EOS [0]{\spacefactor3000\relax}%
\providecommand \BibitemShut  [1]{\csname bibitem#1\endcsname}%
\let\auto@bib@innerbib\@empty
\bibitem [{\citenamefont {Cao}\ \emph {et~al.}(2018{\natexlab{a}})\citenamefont
  {Cao}, \citenamefont {Fatemi}, \citenamefont {Fang}, \citenamefont
  {Watanabe}, \citenamefont {Taniguchi}, \citenamefont {Kaxiras},\ and\
  \citenamefont {Jarillo-Herrero}}]{Cao2018_sc}%
  \BibitemOpen
  \bibfield  {author} {\bibinfo {author} {\bibfnamefont {Y.}~\bibnamefont
  {Cao}}, \bibinfo {author} {\bibfnamefont {V.}~\bibnamefont {Fatemi}},
  \bibinfo {author} {\bibfnamefont {S.}~\bibnamefont {Fang}}, \bibinfo {author}
  {\bibfnamefont {K.}~\bibnamefont {Watanabe}}, \bibinfo {author}
  {\bibfnamefont {T.}~\bibnamefont {Taniguchi}}, \bibinfo {author}
  {\bibfnamefont {E.}~\bibnamefont {Kaxiras}},\ and\ \bibinfo {author}
  {\bibfnamefont {P.}~\bibnamefont {Jarillo-Herrero}},\ }\bibfield  {title}
  {\bibinfo {title} {Unconventional superconductivity in magic-angle graphene
  superlattices},\ }\href {https://doi.org/10.1038/nature26160} {\bibfield
  {journal} {\bibinfo  {journal} {Nature}\ }\textbf {\bibinfo {volume} {556}},\
  \bibinfo {pages} {43} (\bibinfo {year} {2018}{\natexlab{a}})}\BibitemShut
  {NoStop}%
\bibitem [{\citenamefont {Cao}\ \emph {et~al.}(2018{\natexlab{b}})\citenamefont
  {Cao}, \citenamefont {Fatemi}, \citenamefont {Demir}, \citenamefont {Fang},
  \citenamefont {Tomarken}, \citenamefont {Luo}, \citenamefont
  {Sanchez-Yamagishi}, \citenamefont {Watanabe}, \citenamefont {Taniguchi},
  \citenamefont {Kaxiras}, \citenamefont {Ashoori},\ and\ \citenamefont
  {Jarillo-Herrero}}]{Cao2018_corr}%
  \BibitemOpen
  \bibfield  {author} {\bibinfo {author} {\bibfnamefont {Y.}~\bibnamefont
  {Cao}}, \bibinfo {author} {\bibfnamefont {V.}~\bibnamefont {Fatemi}},
  \bibinfo {author} {\bibfnamefont {A.}~\bibnamefont {Demir}}, \bibinfo
  {author} {\bibfnamefont {S.}~\bibnamefont {Fang}}, \bibinfo {author}
  {\bibfnamefont {S.~L.}\ \bibnamefont {Tomarken}}, \bibinfo {author}
  {\bibfnamefont {J.~Y.}\ \bibnamefont {Luo}}, \bibinfo {author} {\bibfnamefont
  {J.~D.}\ \bibnamefont {Sanchez-Yamagishi}}, \bibinfo {author} {\bibfnamefont
  {K.}~\bibnamefont {Watanabe}}, \bibinfo {author} {\bibfnamefont
  {T.}~\bibnamefont {Taniguchi}}, \bibinfo {author} {\bibfnamefont
  {E.}~\bibnamefont {Kaxiras}}, \bibinfo {author} {\bibfnamefont {R.~C.}\
  \bibnamefont {Ashoori}},\ and\ \bibinfo {author} {\bibfnamefont
  {P.}~\bibnamefont {Jarillo-Herrero}},\ }\bibfield  {title} {\bibinfo {title}
  {Correlated insulator behaviour at half-filling in magic-angle graphene
  superlattices},\ }\href {https://doi.org/10.1038/nature26154} {\bibfield
  {journal} {\bibinfo  {journal} {Nature}\ }\textbf {\bibinfo {volume} {556}},\
  \bibinfo {pages} {80} (\bibinfo {year} {2018}{\natexlab{b}})}\BibitemShut
  {NoStop}%
\bibitem [{\citenamefont {Dean}\ \emph {et~al.}(2013)\citenamefont {Dean},
  \citenamefont {Wang}, \citenamefont {Maher}, \citenamefont {Forsythe},
  \citenamefont {Ghahari}, \citenamefont {Gao}, \citenamefont {Katoch},
  \citenamefont {Ishigami}, \citenamefont {Moon}, \citenamefont {Koshino},
  \citenamefont {Taniguchi}, \citenamefont {Watanabe}, \citenamefont {Shepard},
  \citenamefont {Hone},\ and\ \citenamefont {Kim}}]{Dean2013}%
  \BibitemOpen
  \bibfield  {author} {\bibinfo {author} {\bibfnamefont {C.~R.}\ \bibnamefont
  {Dean}}, \bibinfo {author} {\bibfnamefont {L.}~\bibnamefont {Wang}}, \bibinfo
  {author} {\bibfnamefont {P.}~\bibnamefont {Maher}}, \bibinfo {author}
  {\bibfnamefont {C.}~\bibnamefont {Forsythe}}, \bibinfo {author}
  {\bibfnamefont {F.}~\bibnamefont {Ghahari}}, \bibinfo {author} {\bibfnamefont
  {Y.}~\bibnamefont {Gao}}, \bibinfo {author} {\bibfnamefont {J.}~\bibnamefont
  {Katoch}}, \bibinfo {author} {\bibfnamefont {M.}~\bibnamefont {Ishigami}},
  \bibinfo {author} {\bibfnamefont {P.}~\bibnamefont {Moon}}, \bibinfo {author}
  {\bibfnamefont {M.}~\bibnamefont {Koshino}}, \bibinfo {author} {\bibfnamefont
  {T.}~\bibnamefont {Taniguchi}}, \bibinfo {author} {\bibfnamefont
  {K.}~\bibnamefont {Watanabe}}, \bibinfo {author} {\bibfnamefont {K.~L.}\
  \bibnamefont {Shepard}}, \bibinfo {author} {\bibfnamefont {J.}~\bibnamefont
  {Hone}},\ and\ \bibinfo {author} {\bibfnamefont {P.}~\bibnamefont {Kim}},\
  }\bibfield  {title} {\bibinfo {title} {Hofstadter's butterfly and the fractal
  quantum hall effect in moir{\'{e}} superlattices},\ }\href
  {https://doi.org/10.1038/nature12186} {\bibfield  {journal} {\bibinfo
  {journal} {Nature}\ }\textbf {\bibinfo {volume} {497}},\ \bibinfo {pages}
  {598} (\bibinfo {year} {2013})}\BibitemShut {NoStop}%
\bibitem [{\citenamefont {Carr}\ \emph {et~al.}(2017)\citenamefont {Carr},
  \citenamefont {Massatt}, \citenamefont {Fang}, \citenamefont {Cazeaux},
  \citenamefont {Luskin},\ and\ \citenamefont
  {Kaxiras}}]{Carr2017_twistronics}%
  \BibitemOpen
  \bibfield  {author} {\bibinfo {author} {\bibfnamefont {S.}~\bibnamefont
  {Carr}}, \bibinfo {author} {\bibfnamefont {D.}~\bibnamefont {Massatt}},
  \bibinfo {author} {\bibfnamefont {S.}~\bibnamefont {Fang}}, \bibinfo {author}
  {\bibfnamefont {P.}~\bibnamefont {Cazeaux}}, \bibinfo {author} {\bibfnamefont
  {M.}~\bibnamefont {Luskin}},\ and\ \bibinfo {author} {\bibfnamefont
  {E.}~\bibnamefont {Kaxiras}},\ }\bibfield  {title} {\bibinfo {title}
  {Twistronics: Manipulating the electronic properties of two-dimensional
  layered structures through their twist angle},\ }\href
  {https://doi.org/10.1103/PhysRevB.95.075420} {\bibfield  {journal} {\bibinfo
  {journal} {Physical Review B}\ }\textbf {\bibinfo {volume} {95}},\ \bibinfo
  {pages} {075420} (\bibinfo {year} {2017})}\BibitemShut {NoStop}%
\bibitem [{\citenamefont {Tritsaris}\ \emph {et~al.}(2021)\citenamefont
  {Tritsaris}, \citenamefont {Carr},\ and\ \citenamefont
  {Schleder}}]{Tritsaris2021}%
  \BibitemOpen
  \bibfield  {author} {\bibinfo {author} {\bibfnamefont {G.~A.}\ \bibnamefont
  {Tritsaris}}, \bibinfo {author} {\bibfnamefont {S.}~\bibnamefont {Carr}},\
  and\ \bibinfo {author} {\bibfnamefont {G.~R.}\ \bibnamefont {Schleder}},\
  }\bibfield  {title} {\bibinfo {title} {Computational design of moir{\'{e}}
  assemblies aided by artificial intelligence},\ }\href
  {https://doi.org/10.1063/5.0044511} {\bibfield  {journal} {\bibinfo
  {journal} {Applied Physics Reviews}\ }\textbf {\bibinfo {volume} {8}},\
  \bibinfo {pages} {031401} (\bibinfo {year} {2021})}\BibitemShut {NoStop}%
\bibitem [{\citenamefont {Andrei}\ \emph {et~al.}(2021)\citenamefont {Andrei},
  \citenamefont {Efetov}, \citenamefont {Jarillo-Herrero}, \citenamefont
  {MacDonald}, \citenamefont {Mak}, \citenamefont {Senthil}, \citenamefont
  {Tutuc}, \citenamefont {Yazdani},\ and\ \citenamefont {Young}}]{Andrei2021}%
  \BibitemOpen
  \bibfield  {author} {\bibinfo {author} {\bibfnamefont {E.~Y.}\ \bibnamefont
  {Andrei}}, \bibinfo {author} {\bibfnamefont {D.~K.}\ \bibnamefont {Efetov}},
  \bibinfo {author} {\bibfnamefont {P.}~\bibnamefont {Jarillo-Herrero}},
  \bibinfo {author} {\bibfnamefont {A.~H.}\ \bibnamefont {MacDonald}}, \bibinfo
  {author} {\bibfnamefont {K.~F.}\ \bibnamefont {Mak}}, \bibinfo {author}
  {\bibfnamefont {T.}~\bibnamefont {Senthil}}, \bibinfo {author} {\bibfnamefont
  {E.}~\bibnamefont {Tutuc}}, \bibinfo {author} {\bibfnamefont
  {A.}~\bibnamefont {Yazdani}},\ and\ \bibinfo {author} {\bibfnamefont {A.~F.}\
  \bibnamefont {Young}},\ }\bibfield  {title} {\bibinfo {title} {The marvels of
  moir{\'e} materials},\ }\href {https://doi.org/10.1038/s41578-021-00284-1}
  {\bibfield  {journal} {\bibinfo  {journal} {Nature Reviews Materials}\
  }\textbf {\bibinfo {volume} {6}},\ \bibinfo {pages} {201} (\bibinfo {year}
  {2021})}\BibitemShut {NoStop}%
\bibitem [{\citenamefont {Angeli}\ \emph {et~al.}(2022)\citenamefont {Angeli},
  \citenamefont {Schleder},\ and\ \citenamefont {Kaxiras}}]{Angeli2022}%
  \BibitemOpen
  \bibfield  {author} {\bibinfo {author} {\bibfnamefont {M.}~\bibnamefont
  {Angeli}}, \bibinfo {author} {\bibfnamefont {G.~R.}\ \bibnamefont
  {Schleder}},\ and\ \bibinfo {author} {\bibfnamefont {E.}~\bibnamefont
  {Kaxiras}},\ }\bibfield  {title} {\bibinfo {title} {Twistronics of janus
  transition metal dichalcogenide bilayers},\ }\href
  {https://doi.org/10.1103/physrevb.106.235159} {\bibfield  {journal} {\bibinfo
   {journal} {Physical Review B}\ }\textbf {\bibinfo {volume} {106}},\ \bibinfo
  {pages} {235159} (\bibinfo {year} {2022})}\BibitemShut {NoStop}%
\bibitem [{\citenamefont {Kennes}\ \emph {et~al.}(2021)\citenamefont {Kennes},
  \citenamefont {Claassen}, \citenamefont {Xian}, \citenamefont {Georges},
  \citenamefont {Millis}, \citenamefont {Hone}, \citenamefont {Dean},
  \citenamefont {Basov}, \citenamefont {Pasupathy},\ and\ \citenamefont
  {Rubio}}]{Kennes2021}%
  \BibitemOpen
  \bibfield  {author} {\bibinfo {author} {\bibfnamefont {D.~M.}\ \bibnamefont
  {Kennes}}, \bibinfo {author} {\bibfnamefont {M.}~\bibnamefont {Claassen}},
  \bibinfo {author} {\bibfnamefont {L.}~\bibnamefont {Xian}}, \bibinfo {author}
  {\bibfnamefont {A.}~\bibnamefont {Georges}}, \bibinfo {author} {\bibfnamefont
  {A.~J.}\ \bibnamefont {Millis}}, \bibinfo {author} {\bibfnamefont
  {J.}~\bibnamefont {Hone}}, \bibinfo {author} {\bibfnamefont {C.~R.}\
  \bibnamefont {Dean}}, \bibinfo {author} {\bibfnamefont {D.~N.}\ \bibnamefont
  {Basov}}, \bibinfo {author} {\bibfnamefont {A.~N.}\ \bibnamefont
  {Pasupathy}},\ and\ \bibinfo {author} {\bibfnamefont {A.}~\bibnamefont
  {Rubio}},\ }\bibfield  {title} {\bibinfo {title} {Moir{\'{e}}
  heterostructures as a condensed-matter quantum simulator},\ }\href
  {https://doi.org/10.1038/s41567-020-01154-3} {\bibfield  {journal} {\bibinfo
  {journal} {Nature Physics}\ }\textbf {\bibinfo {volume} {17}},\ \bibinfo
  {pages} {155} (\bibinfo {year} {2021})}\BibitemShut {NoStop}%
\bibitem [{\citenamefont {Waters}\ \emph {et~al.}(2020)\citenamefont {Waters},
  \citenamefont {Nie}, \citenamefont {L\"{u}pke}, \citenamefont {Pan},
  \citenamefont {F\"{o}lsch}, \citenamefont {Lin}, \citenamefont {Jariwala},
  \citenamefont {Zhang}, \citenamefont {Wang}, \citenamefont {Lv},
  \citenamefont {Cho}, \citenamefont {Xiao}, \citenamefont {Robinson},\ and\
  \citenamefont {Feenstra}}]{Waters2020}%
  \BibitemOpen
  \bibfield  {author} {\bibinfo {author} {\bibfnamefont {D.}~\bibnamefont
  {Waters}}, \bibinfo {author} {\bibfnamefont {Y.}~\bibnamefont {Nie}},
  \bibinfo {author} {\bibfnamefont {F.}~\bibnamefont {L\"{u}pke}}, \bibinfo
  {author} {\bibfnamefont {Y.}~\bibnamefont {Pan}}, \bibinfo {author}
  {\bibfnamefont {S.}~\bibnamefont {F\"{o}lsch}}, \bibinfo {author}
  {\bibfnamefont {Y.-C.}\ \bibnamefont {Lin}}, \bibinfo {author} {\bibfnamefont
  {B.}~\bibnamefont {Jariwala}}, \bibinfo {author} {\bibfnamefont
  {K.}~\bibnamefont {Zhang}}, \bibinfo {author} {\bibfnamefont
  {C.}~\bibnamefont {Wang}}, \bibinfo {author} {\bibfnamefont {H.}~\bibnamefont
  {Lv}}, \bibinfo {author} {\bibfnamefont {K.}~\bibnamefont {Cho}}, \bibinfo
  {author} {\bibfnamefont {D.}~\bibnamefont {Xiao}}, \bibinfo {author}
  {\bibfnamefont {J.~A.}\ \bibnamefont {Robinson}},\ and\ \bibinfo {author}
  {\bibfnamefont {R.~M.}\ \bibnamefont {Feenstra}},\ }\bibfield  {title}
  {\bibinfo {title} {Flat bands and mechanical deformation effects in the
  moir{\'{e}} superlattice of {MoS}$_2$-{WSe}$_2$ heterobilayers},\ }\href
  {https://doi.org/10.1021/acsnano.0c03414} {\bibfield  {journal} {\bibinfo
  {journal} {{ACS} Nano}\ }\textbf {\bibinfo {volume} {14}},\ \bibinfo {pages}
  {7564} (\bibinfo {year} {2020})}\BibitemShut {NoStop}%
\bibitem [{\citenamefont {Salvato}\ \emph {et~al.}(2022)\citenamefont
  {Salvato}, \citenamefont {Crescenzi}, \citenamefont {Scagliotti},
  \citenamefont {Castrucci}, \citenamefont {Boninelli}, \citenamefont {Caruso},
  \citenamefont {Liu}, \citenamefont {Mikkelsen}, \citenamefont {Timm},
  \citenamefont {Nahas}, \citenamefont {Black-Schaffer}, \citenamefont
  {Kunakova}, \citenamefont {Andzane}, \citenamefont {Erts}, \citenamefont
  {Bauch},\ and\ \citenamefont {Lombardi}}]{Salvato2022}%
  \BibitemOpen
  \bibfield  {author} {\bibinfo {author} {\bibfnamefont {M.}~\bibnamefont
  {Salvato}}, \bibinfo {author} {\bibfnamefont {M.~D.}\ \bibnamefont
  {Crescenzi}}, \bibinfo {author} {\bibfnamefont {M.}~\bibnamefont
  {Scagliotti}}, \bibinfo {author} {\bibfnamefont {P.}~\bibnamefont
  {Castrucci}}, \bibinfo {author} {\bibfnamefont {S.}~\bibnamefont
  {Boninelli}}, \bibinfo {author} {\bibfnamefont {G.~M.}\ \bibnamefont
  {Caruso}}, \bibinfo {author} {\bibfnamefont {Y.}~\bibnamefont {Liu}},
  \bibinfo {author} {\bibfnamefont {A.}~\bibnamefont {Mikkelsen}}, \bibinfo
  {author} {\bibfnamefont {R.}~\bibnamefont {Timm}}, \bibinfo {author}
  {\bibfnamefont {S.}~\bibnamefont {Nahas}}, \bibinfo {author} {\bibfnamefont
  {A.}~\bibnamefont {Black-Schaffer}}, \bibinfo {author} {\bibfnamefont
  {G.}~\bibnamefont {Kunakova}}, \bibinfo {author} {\bibfnamefont
  {J.}~\bibnamefont {Andzane}}, \bibinfo {author} {\bibfnamefont
  {D.}~\bibnamefont {Erts}}, \bibinfo {author} {\bibfnamefont {T.}~\bibnamefont
  {Bauch}},\ and\ \bibinfo {author} {\bibfnamefont {F.}~\bibnamefont
  {Lombardi}},\ }\bibfield  {title} {\bibinfo {title} {Nanometric moir{\'{e}}
  stripes on the surface of {Bi}$_2${Se}$_3$ topological insulator},\ }\href
  {https://doi.org/10.1021/acsnano.2c02515} {\bibfield  {journal} {\bibinfo
  {journal} {{ACS} Nano}\ }\textbf {\bibinfo {volume} {16}},\ \bibinfo {pages}
  {13860} (\bibinfo {year} {2022})}\BibitemShut {NoStop}%
\bibitem [{\citenamefont {Rakib}\ \emph {et~al.}(2022)\citenamefont {Rakib},
  \citenamefont {Pochet}, \citenamefont {Ertekin},\ and\ \citenamefont
  {Johnson}}]{Rakib2022}%
  \BibitemOpen
  \bibfield  {author} {\bibinfo {author} {\bibfnamefont {T.}~\bibnamefont
  {Rakib}}, \bibinfo {author} {\bibfnamefont {P.}~\bibnamefont {Pochet}},
  \bibinfo {author} {\bibfnamefont {E.}~\bibnamefont {Ertekin}},\ and\ \bibinfo
  {author} {\bibfnamefont {H.~T.}\ \bibnamefont {Johnson}},\ }\bibfield
  {title} {\bibinfo {title} {Moir{\'{e}} engineering in van der waals
  heterostructures},\ }\href {https://doi.org/10.1063/5.0105405} {\bibfield
  {journal} {\bibinfo  {journal} {Journal of Applied Physics}\ }\textbf
  {\bibinfo {volume} {132}},\ \bibinfo {pages} {120901} (\bibinfo {year}
  {2022})}\BibitemShut {NoStop}%
\bibitem [{\citenamefont {Li}\ \emph {et~al.}(2023)\citenamefont {Li},
  \citenamefont {Yuan}, \citenamefont {Guo}, \citenamefont {Lou}, \citenamefont
  {Cui}, \citenamefont {Mei}, \citenamefont {Petek}, \citenamefont {Cao},
  \citenamefont {Ji},\ and\ \citenamefont {Feng}}]{Li2023}%
  \BibitemOpen
  \bibfield  {author} {\bibinfo {author} {\bibfnamefont {Y.}~\bibnamefont
  {Li}}, \bibinfo {author} {\bibfnamefont {Q.}~\bibnamefont {Yuan}}, \bibinfo
  {author} {\bibfnamefont {D.}~\bibnamefont {Guo}}, \bibinfo {author}
  {\bibfnamefont {C.}~\bibnamefont {Lou}}, \bibinfo {author} {\bibfnamefont
  {X.}~\bibnamefont {Cui}}, \bibinfo {author} {\bibfnamefont {G.}~\bibnamefont
  {Mei}}, \bibinfo {author} {\bibfnamefont {H.}~\bibnamefont {Petek}}, \bibinfo
  {author} {\bibfnamefont {L.}~\bibnamefont {Cao}}, \bibinfo {author}
  {\bibfnamefont {W.}~\bibnamefont {Ji}},\ and\ \bibinfo {author}
  {\bibfnamefont {M.}~\bibnamefont {Feng}},\ }\bibfield  {title} {\bibinfo
  {title} {1d electronic flat bands in untwisted moir{\'{e}} superlattices},\
  }\bibfield  {journal} {\bibinfo  {journal} {Advanced Materials}\ }\href
  {https://doi.org/10.1002/adma.202300572} {10.1002/adma.202300572} (\bibinfo
  {year} {2023})\BibitemShut {NoStop}%
\bibitem [{\citenamefont {Bonnet}\ \emph {et~al.}(2016)\citenamefont {Bonnet},
  \citenamefont {Lherbier}, \citenamefont {Barraud}, \citenamefont {Rocca},
  \citenamefont {Lafarge},\ and\ \citenamefont {Charlier}}]{Bonnet2016}%
  \BibitemOpen
  \bibfield  {author} {\bibinfo {author} {\bibfnamefont {R.}~\bibnamefont
  {Bonnet}}, \bibinfo {author} {\bibfnamefont {A.}~\bibnamefont {Lherbier}},
  \bibinfo {author} {\bibfnamefont {C.}~\bibnamefont {Barraud}}, \bibinfo
  {author} {\bibfnamefont {M.~L.~D.}\ \bibnamefont {Rocca}}, \bibinfo {author}
  {\bibfnamefont {P.}~\bibnamefont {Lafarge}},\ and\ \bibinfo {author}
  {\bibfnamefont {J.-C.}\ \bibnamefont {Charlier}},\ }\bibfield  {title}
  {\bibinfo {title} {Charge transport through one-dimensional moir{\'{e}}
  crystals},\ }\href {https://doi.org/10.1038/srep19701} {\bibfield  {journal}
  {\bibinfo  {journal} {Scientific Reports}\ }\textbf {\bibinfo {volume} {6}},\
  \bibinfo {pages} {19701} (\bibinfo {year} {2016})}\BibitemShut {NoStop}%
\bibitem [{\citenamefont {Arroyo-Gasc{\'{o}}n}\ \emph
  {et~al.}(2020)\citenamefont {Arroyo-Gasc{\'{o}}n}, \citenamefont
  {Fern{\'{a}}ndez-Perea}, \citenamefont {Morell}, \citenamefont {Cabrillo},\
  and\ \citenamefont {Chico}}]{ArroyoGascn2020}%
  \BibitemOpen
  \bibfield  {author} {\bibinfo {author} {\bibfnamefont {O.}~\bibnamefont
  {Arroyo-Gasc{\'{o}}n}}, \bibinfo {author} {\bibfnamefont {R.}~\bibnamefont
  {Fern{\'{a}}ndez-Perea}}, \bibinfo {author} {\bibfnamefont {E.~S.}\
  \bibnamefont {Morell}}, \bibinfo {author} {\bibfnamefont {C.}~\bibnamefont
  {Cabrillo}},\ and\ \bibinfo {author} {\bibfnamefont {L.}~\bibnamefont
  {Chico}},\ }\bibfield  {title} {\bibinfo {title} {One-dimensional moir{\'{e}}
  superlattices and flat bands in collapsed chiral carbon nanotubes},\ }\href
  {https://doi.org/10.1021/acs.nanolett.0c03091} {\bibfield  {journal}
  {\bibinfo  {journal} {Nano Letters}\ }\textbf {\bibinfo {volume} {20}},\
  \bibinfo {pages} {7588} (\bibinfo {year} {2020})}\BibitemShut {NoStop}%
\bibitem [{\citenamefont {Li}\ \emph {et~al.}(2021)\citenamefont {Li},
  \citenamefont {Dietrich}, \citenamefont {Forsythe}, \citenamefont
  {Taniguchi}, \citenamefont {Watanabe}, \citenamefont {Moon},\ and\
  \citenamefont {Dean}}]{Li2021}%
  \BibitemOpen
  \bibfield  {author} {\bibinfo {author} {\bibfnamefont {Y.}~\bibnamefont
  {Li}}, \bibinfo {author} {\bibfnamefont {S.}~\bibnamefont {Dietrich}},
  \bibinfo {author} {\bibfnamefont {C.}~\bibnamefont {Forsythe}}, \bibinfo
  {author} {\bibfnamefont {T.}~\bibnamefont {Taniguchi}}, \bibinfo {author}
  {\bibfnamefont {K.}~\bibnamefont {Watanabe}}, \bibinfo {author}
  {\bibfnamefont {P.}~\bibnamefont {Moon}},\ and\ \bibinfo {author}
  {\bibfnamefont {C.~R.}\ \bibnamefont {Dean}},\ }\bibfield  {title} {\bibinfo
  {title} {Anisotropic band flattening in graphene with one-dimensional
  superlattices},\ }\href {https://doi.org/10.1038/s41565-021-00849-9}
  {\bibfield  {journal} {\bibinfo  {journal} {Nature Nanotechnology}\ }\textbf
  {\bibinfo {volume} {16}},\ \bibinfo {pages} {525} (\bibinfo {year}
  {2021})}\BibitemShut {NoStop}%
\bibitem [{\citenamefont {Gupta}\ \emph {et~al.}(2022)\citenamefont {Gupta},
  \citenamefont {Yu},\ and\ \citenamefont {Yakobson}}]{Gupta2022}%
  \BibitemOpen
  \bibfield  {author} {\bibinfo {author} {\bibfnamefont {S.}~\bibnamefont
  {Gupta}}, \bibinfo {author} {\bibfnamefont {H.}~\bibnamefont {Yu}},\ and\
  \bibinfo {author} {\bibfnamefont {B.~I.}\ \bibnamefont {Yakobson}},\
  }\bibfield  {title} {\bibinfo {title} {Designing 1d correlated-electron
  states by non-euclidean topography of 2d monolayers},\ }\href
  {https://doi.org/10.1038/s41467-022-30818-2} {\bibfield  {journal} {\bibinfo
  {journal} {Nature Communications}\ }\textbf {\bibinfo {volume} {13}},\
  \bibinfo {pages} {3103} (\bibinfo {year} {2022})}\BibitemShut {NoStop}%
\bibitem [{\citenamefont {Yang}\ and\ \citenamefont {Zhang}(2023)}]{Yang2023}%
  \BibitemOpen
  \bibfield  {author} {\bibinfo {author} {\bibfnamefont {X.}~\bibnamefont
  {Yang}}\ and\ \bibinfo {author} {\bibfnamefont {B.}~\bibnamefont {Zhang}},\
  }\bibfield  {title} {\bibinfo {title} {Heterostrain and temperature-tuned
  twist between graphene/h-{BN} bilayers},\ }\bibfield  {journal} {\bibinfo
  {journal} {Scientific Reports}\ }\textbf {\bibinfo {volume} {13}},\ \href
  {https://doi.org/10.1038/s41598-023-31233-3} {10.1038/s41598-023-31233-3}
  (\bibinfo {year} {2023})\BibitemShut {NoStop}%
\bibitem [{\citenamefont {Cosma}\ \emph {et~al.}(2014)\citenamefont {Cosma},
  \citenamefont {Wallbank}, \citenamefont {Cheianov},\ and\ \citenamefont
  {Fal{\textquotesingle}ko}}]{Cosma2014}%
  \BibitemOpen
  \bibfield  {author} {\bibinfo {author} {\bibfnamefont {D.~A.}\ \bibnamefont
  {Cosma}}, \bibinfo {author} {\bibfnamefont {J.~R.}\ \bibnamefont {Wallbank}},
  \bibinfo {author} {\bibfnamefont {V.}~\bibnamefont {Cheianov}},\ and\
  \bibinfo {author} {\bibfnamefont {V.~I.}\ \bibnamefont
  {Fal{\textquotesingle}ko}},\ }\bibfield  {title} {\bibinfo {title}
  {Moir{\'{e}} pattern as a magnifying glass for strain and dislocations in van
  der waals heterostructures},\ }\href {https://doi.org/10.1039/c4fd00146j}
  {\bibfield  {journal} {\bibinfo  {journal} {Faraday Discuss.}\ }\textbf
  {\bibinfo {volume} {173}},\ \bibinfo {pages} {137} (\bibinfo {year}
  {2014})}\BibitemShut {NoStop}%
\bibitem [{\citenamefont {Bi}\ \emph {et~al.}(2019)\citenamefont {Bi},
  \citenamefont {Yuan},\ and\ \citenamefont {Fu}}]{Bi2019}%
  \BibitemOpen
  \bibfield  {author} {\bibinfo {author} {\bibfnamefont {Z.}~\bibnamefont
  {Bi}}, \bibinfo {author} {\bibfnamefont {N.~F.~Q.}\ \bibnamefont {Yuan}},\
  and\ \bibinfo {author} {\bibfnamefont {L.}~\bibnamefont {Fu}},\ }\bibfield
  {title} {\bibinfo {title} {Designing flat bands by strain},\ }\href
  {https://doi.org/10.1103/physrevb.100.035448} {\bibfield  {journal} {\bibinfo
   {journal} {Physical Review B}\ }\textbf {\bibinfo {volume} {100}},\ \bibinfo
  {pages} {035448} (\bibinfo {year} {2019})}\BibitemShut {NoStop}%
\bibitem [{\citenamefont {Zhu}\ and\ \citenamefont {Johnson}(2018)}]{Zhu2018}%
  \BibitemOpen
  \bibfield  {author} {\bibinfo {author} {\bibfnamefont {S.}~\bibnamefont
  {Zhu}}\ and\ \bibinfo {author} {\bibfnamefont {H.~T.}\ \bibnamefont
  {Johnson}},\ }\bibfield  {title} {\bibinfo {title} {Moir{\'{e}}-templated
  strain patterning in transition-metal dichalcogenides and application in
  twisted bilayer {MoS}$_2$},\ }\href {https://doi.org/10.1039/c8nr06269b}
  {\bibfield  {journal} {\bibinfo  {journal} {Nanoscale}\ }\textbf {\bibinfo
  {volume} {10}},\ \bibinfo {pages} {20689} (\bibinfo {year}
  {2018})}\BibitemShut {NoStop}%
\bibitem [{\citenamefont {K\"{o}gl}\ \emph {et~al.}(2023)\citenamefont
  {K\"{o}gl}, \citenamefont {Soubelet}, \citenamefont {Brotons-Gisbert},
  \citenamefont {Stier}, \citenamefont {Gerardot},\ and\ \citenamefont
  {Finley}}]{Kogl2023}%
  \BibitemOpen
  \bibfield  {author} {\bibinfo {author} {\bibfnamefont {M.}~\bibnamefont
  {K\"{o}gl}}, \bibinfo {author} {\bibfnamefont {P.}~\bibnamefont {Soubelet}},
  \bibinfo {author} {\bibfnamefont {M.}~\bibnamefont {Brotons-Gisbert}},
  \bibinfo {author} {\bibfnamefont {A.~V.}\ \bibnamefont {Stier}}, \bibinfo
  {author} {\bibfnamefont {B.~D.}\ \bibnamefont {Gerardot}},\ and\ \bibinfo
  {author} {\bibfnamefont {J.~J.}\ \bibnamefont {Finley}},\ }\bibfield  {title}
  {\bibinfo {title} {Moir{\'{e}} straintronics: a universal platform for
  reconfigurable quantum materials},\ }\bibfield  {journal} {\bibinfo
  {journal} {npj 2D Materials and Applications}\ }\textbf {\bibinfo {volume}
  {7}},\ \href {https://doi.org/10.1038/s41699-023-00382-4}
  {10.1038/s41699-023-00382-4} (\bibinfo {year} {2023})\BibitemShut {NoStop}%
\bibitem [{\citenamefont {Bai}\ \emph {et~al.}(2020)\citenamefont {Bai},
  \citenamefont {Zhou}, \citenamefont {Wang}, \citenamefont {Wu}, \citenamefont
  {McGilly}, \citenamefont {Halbertal}, \citenamefont {Lo}, \citenamefont
  {Liu}, \citenamefont {Ardelean}, \citenamefont {Rivera}, \citenamefont
  {Finney}, \citenamefont {Yang}, \citenamefont {Basov}, \citenamefont {Yao},
  \citenamefont {Xu}, \citenamefont {Hone}, \citenamefont {Pasupathy},\ and\
  \citenamefont {Zhu}}]{Bai2020}%
  \BibitemOpen
  \bibfield  {author} {\bibinfo {author} {\bibfnamefont {Y.}~\bibnamefont
  {Bai}}, \bibinfo {author} {\bibfnamefont {L.}~\bibnamefont {Zhou}}, \bibinfo
  {author} {\bibfnamefont {J.}~\bibnamefont {Wang}}, \bibinfo {author}
  {\bibfnamefont {W.}~\bibnamefont {Wu}}, \bibinfo {author} {\bibfnamefont
  {L.~J.}\ \bibnamefont {McGilly}}, \bibinfo {author} {\bibfnamefont
  {D.}~\bibnamefont {Halbertal}}, \bibinfo {author} {\bibfnamefont {C.~F.~B.}\
  \bibnamefont {Lo}}, \bibinfo {author} {\bibfnamefont {F.}~\bibnamefont
  {Liu}}, \bibinfo {author} {\bibfnamefont {J.}~\bibnamefont {Ardelean}},
  \bibinfo {author} {\bibfnamefont {P.}~\bibnamefont {Rivera}}, \bibinfo
  {author} {\bibfnamefont {N.~R.}\ \bibnamefont {Finney}}, \bibinfo {author}
  {\bibfnamefont {X.-C.}\ \bibnamefont {Yang}}, \bibinfo {author}
  {\bibfnamefont {D.~N.}\ \bibnamefont {Basov}}, \bibinfo {author}
  {\bibfnamefont {W.}~\bibnamefont {Yao}}, \bibinfo {author} {\bibfnamefont
  {X.}~\bibnamefont {Xu}}, \bibinfo {author} {\bibfnamefont {J.}~\bibnamefont
  {Hone}}, \bibinfo {author} {\bibfnamefont {A.~N.}\ \bibnamefont
  {Pasupathy}},\ and\ \bibinfo {author} {\bibfnamefont {X.-Y.}\ \bibnamefont
  {Zhu}},\ }\bibfield  {title} {\bibinfo {title} {Excitons in strain-induced
  one-dimensional moir{\'{e}} potentials at transition metal dichalcogenide
  heterojunctions},\ }\href {https://doi.org/10.1038/s41563-020-0730-8}
  {\bibfield  {journal} {\bibinfo  {journal} {Nature Materials}\ }\textbf
  {\bibinfo {volume} {19}},\ \bibinfo {pages} {1068} (\bibinfo {year}
  {2020})}\BibitemShut {NoStop}%
\bibitem [{\citenamefont {Gao}\ \emph {et~al.}(2021)\citenamefont {Gao},
  \citenamefont {Sun}, \citenamefont {Kang}, \citenamefont {Wang},
  \citenamefont {Wang},\ and\ \citenamefont {Nam}}]{Gao2021}%
  \BibitemOpen
  \bibfield  {author} {\bibinfo {author} {\bibfnamefont {X.}~\bibnamefont
  {Gao}}, \bibinfo {author} {\bibfnamefont {H.}~\bibnamefont {Sun}}, \bibinfo
  {author} {\bibfnamefont {D.-H.}\ \bibnamefont {Kang}}, \bibinfo {author}
  {\bibfnamefont {C.}~\bibnamefont {Wang}}, \bibinfo {author} {\bibfnamefont
  {Q.~J.}\ \bibnamefont {Wang}},\ and\ \bibinfo {author} {\bibfnamefont
  {D.}~\bibnamefont {Nam}},\ }\bibfield  {title} {\bibinfo {title}
  {Heterostrain-enabled dynamically tunable moir{\'{e}} superlattice in twisted
  bilayer graphene},\ }\href {https://doi.org/10.1038/s41598-021-00757-x}
  {\bibfield  {journal} {\bibinfo  {journal} {Scientific Reports}\ }\textbf
  {\bibinfo {volume} {11}},\ \bibinfo {pages} {21402} (\bibinfo {year}
  {2021})}\BibitemShut {NoStop}%
\bibitem [{\citenamefont {Mesple}\ \emph {et~al.}(2021)\citenamefont {Mesple},
  \citenamefont {Missaoui}, \citenamefont {Cea}, \citenamefont {Huder},
  \citenamefont {Guinea}, \citenamefont {de~Laissardi{\`{e}}re}, \citenamefont
  {Chapelier},\ and\ \citenamefont {Renard}}]{Mesple2021}%
  \BibitemOpen
  \bibfield  {author} {\bibinfo {author} {\bibfnamefont {F.}~\bibnamefont
  {Mesple}}, \bibinfo {author} {\bibfnamefont {A.}~\bibnamefont {Missaoui}},
  \bibinfo {author} {\bibfnamefont {T.}~\bibnamefont {Cea}}, \bibinfo {author}
  {\bibfnamefont {L.}~\bibnamefont {Huder}}, \bibinfo {author} {\bibfnamefont
  {F.}~\bibnamefont {Guinea}}, \bibinfo {author} {\bibfnamefont {G.~T.}\
  \bibnamefont {de~Laissardi{\`{e}}re}}, \bibinfo {author} {\bibfnamefont
  {C.}~\bibnamefont {Chapelier}},\ and\ \bibinfo {author} {\bibfnamefont
  {V.~T.}\ \bibnamefont {Renard}},\ }\bibfield  {title} {\bibinfo {title}
  {Heterostrain determines flat bands in magic-angle twisted graphene layers},\
  }\href {https://doi.org/10.1103/physrevlett.127.126405} {\bibfield  {journal}
  {\bibinfo  {journal} {Physical Review Letters}\ }\textbf {\bibinfo {volume}
  {127}},\ \bibinfo {pages} {126405} (\bibinfo {year} {2021})}\BibitemShut
  {NoStop}%
\bibitem [{\citenamefont {Zhang}\ \emph {et~al.}(2022)\citenamefont {Zhang},
  \citenamefont {Wang}, \citenamefont {Hu}, \citenamefont {Wan}, \citenamefont
  {Zheliuk}, \citenamefont {Liang}, \citenamefont {Peng}, \citenamefont
  {Zeng},\ and\ \citenamefont {Ye}}]{Zhang2022}%
  \BibitemOpen
  \bibfield  {author} {\bibinfo {author} {\bibfnamefont {L.}~\bibnamefont
  {Zhang}}, \bibinfo {author} {\bibfnamefont {Y.}~\bibnamefont {Wang}},
  \bibinfo {author} {\bibfnamefont {R.}~\bibnamefont {Hu}}, \bibinfo {author}
  {\bibfnamefont {P.}~\bibnamefont {Wan}}, \bibinfo {author} {\bibfnamefont
  {O.}~\bibnamefont {Zheliuk}}, \bibinfo {author} {\bibfnamefont
  {M.}~\bibnamefont {Liang}}, \bibinfo {author} {\bibfnamefont
  {X.}~\bibnamefont {Peng}}, \bibinfo {author} {\bibfnamefont {Y.-J.}\
  \bibnamefont {Zeng}},\ and\ \bibinfo {author} {\bibfnamefont
  {J.}~\bibnamefont {Ye}},\ }\bibfield  {title} {\bibinfo {title} {Correlated
  states in strained twisted bilayer graphenes away from the magic angle},\
  }\href {https://doi.org/10.1021/acs.nanolett.1c04400} {\bibfield  {journal}
  {\bibinfo  {journal} {Nano Letters}\ }\textbf {\bibinfo {volume} {22}},\
  \bibinfo {pages} {3204} (\bibinfo {year} {2022})}\BibitemShut {NoStop}%
\bibitem [{\citenamefont {Kapfer}\ \emph {et~al.}(2022)\citenamefont {Kapfer},
  \citenamefont {Jessen}, \citenamefont {Eisele}, \citenamefont {Fu},
  \citenamefont {Danielsen}, \citenamefont {Darlington}, \citenamefont {Moore},
  \citenamefont {Finney}, \citenamefont {Marchese}, \citenamefont {Hsieh},
  \citenamefont {Majchrzak}, \citenamefont {Jiang}, \citenamefont {Biswas},
  \citenamefont {Dudin}, \citenamefont {Avila}, \citenamefont {Watanabe},
  \citenamefont {Taniguchi}, \citenamefont {Ulstrup}, \citenamefont {Bøggild},
  \citenamefont {Schuck}, \citenamefont {Basov}, \citenamefont {Hone},\ and\
  \citenamefont {Dean}}]{Kapfer2023}%
  \BibitemOpen
  \bibfield  {author} {\bibinfo {author} {\bibfnamefont {M.}~\bibnamefont
  {Kapfer}}, \bibinfo {author} {\bibfnamefont {B.~S.}\ \bibnamefont {Jessen}},
  \bibinfo {author} {\bibfnamefont {M.~E.}\ \bibnamefont {Eisele}}, \bibinfo
  {author} {\bibfnamefont {M.}~\bibnamefont {Fu}}, \bibinfo {author}
  {\bibfnamefont {D.~R.}\ \bibnamefont {Danielsen}}, \bibinfo {author}
  {\bibfnamefont {T.~P.}\ \bibnamefont {Darlington}}, \bibinfo {author}
  {\bibfnamefont {S.~L.}\ \bibnamefont {Moore}}, \bibinfo {author}
  {\bibfnamefont {N.~R.}\ \bibnamefont {Finney}}, \bibinfo {author}
  {\bibfnamefont {A.}~\bibnamefont {Marchese}}, \bibinfo {author}
  {\bibfnamefont {V.}~\bibnamefont {Hsieh}}, \bibinfo {author} {\bibfnamefont
  {P.}~\bibnamefont {Majchrzak}}, \bibinfo {author} {\bibfnamefont
  {Z.}~\bibnamefont {Jiang}}, \bibinfo {author} {\bibfnamefont
  {D.}~\bibnamefont {Biswas}}, \bibinfo {author} {\bibfnamefont
  {P.}~\bibnamefont {Dudin}}, \bibinfo {author} {\bibfnamefont
  {J.}~\bibnamefont {Avila}}, \bibinfo {author} {\bibfnamefont
  {K.}~\bibnamefont {Watanabe}}, \bibinfo {author} {\bibfnamefont
  {T.}~\bibnamefont {Taniguchi}}, \bibinfo {author} {\bibfnamefont
  {S.}~\bibnamefont {Ulstrup}}, \bibinfo {author} {\bibfnamefont
  {P.}~\bibnamefont {Bøggild}}, \bibinfo {author} {\bibfnamefont {P.~J.}\
  \bibnamefont {Schuck}}, \bibinfo {author} {\bibfnamefont {D.~N.}\
  \bibnamefont {Basov}}, \bibinfo {author} {\bibfnamefont {J.}~\bibnamefont
  {Hone}},\ and\ \bibinfo {author} {\bibfnamefont {C.~R.}\ \bibnamefont
  {Dean}},\ }\bibfield  {title} {\bibinfo {title} {Programming moiré patterns
  in {2D} materials by bending}\ }\href
  {https://doi.org/10.48550/arxiv.2209.10696} {10.48550/arxiv.2209.10696}
  (\bibinfo {year} {2022})\BibitemShut {NoStop}%
\bibitem [{\citenamefont {Zhu}\ \emph {et~al.}(2012)\citenamefont {Zhu},
  \citenamefont {Low}, \citenamefont {Perebeinos}, \citenamefont {Bol},
  \citenamefont {Zhu}, \citenamefont {Yan}, \citenamefont {Tersoff},\ and\
  \citenamefont {Avouris}}]{Avouris2012}%
  \BibitemOpen
  \bibfield  {author} {\bibinfo {author} {\bibfnamefont {W.}~\bibnamefont
  {Zhu}}, \bibinfo {author} {\bibfnamefont {T.}~\bibnamefont {Low}}, \bibinfo
  {author} {\bibfnamefont {V.}~\bibnamefont {Perebeinos}}, \bibinfo {author}
  {\bibfnamefont {A.~A.}\ \bibnamefont {Bol}}, \bibinfo {author} {\bibfnamefont
  {Y.}~\bibnamefont {Zhu}}, \bibinfo {author} {\bibfnamefont {H.}~\bibnamefont
  {Yan}}, \bibinfo {author} {\bibfnamefont {J.}~\bibnamefont {Tersoff}},\ and\
  \bibinfo {author} {\bibfnamefont {P.}~\bibnamefont {Avouris}},\ }\bibfield
  {title} {\bibinfo {title} {Structure and electronic transport in graphene
  wrinkles},\ }\href {https://doi.org/10.1021/nl300563h} {\bibfield  {journal}
  {\bibinfo  {journal} {Nano Letters}\ }\textbf {\bibinfo {volume} {12}},\
  \bibinfo {pages} {3431} (\bibinfo {year} {2012})}\BibitemShut {NoStop}%
\bibitem [{\citenamefont {Lin}\ \emph {et~al.}(2013)\citenamefont {Lin},
  \citenamefont {Fang}, \citenamefont {Zhou}, \citenamefont {Lupini},
  \citenamefont {Idrobo}, \citenamefont {Kong}, \citenamefont {Pennycook},\
  and\ \citenamefont {Pantelides}}]{Lin2013}%
  \BibitemOpen
  \bibfield  {author} {\bibinfo {author} {\bibfnamefont {J.}~\bibnamefont
  {Lin}}, \bibinfo {author} {\bibfnamefont {W.}~\bibnamefont {Fang}}, \bibinfo
  {author} {\bibfnamefont {W.}~\bibnamefont {Zhou}}, \bibinfo {author}
  {\bibfnamefont {A.~R.}\ \bibnamefont {Lupini}}, \bibinfo {author}
  {\bibfnamefont {J.~C.}\ \bibnamefont {Idrobo}}, \bibinfo {author}
  {\bibfnamefont {J.}~\bibnamefont {Kong}}, \bibinfo {author} {\bibfnamefont
  {S.~J.}\ \bibnamefont {Pennycook}},\ and\ \bibinfo {author} {\bibfnamefont
  {S.~T.}\ \bibnamefont {Pantelides}},\ }\bibfield  {title} {\bibinfo {title}
  {{AC}/{AB} stacking boundaries in bilayer graphene},\ }\href
  {https://doi.org/10.1021/nl4013979} {\bibfield  {journal} {\bibinfo
  {journal} {Nano Letters}\ }\textbf {\bibinfo {volume} {13}},\ \bibinfo
  {pages} {3262} (\bibinfo {year} {2013})}\BibitemShut {NoStop}%
\bibitem [{\citenamefont {Guan}\ and\ \citenamefont
  {Yazyev}(2022)}]{Yazyev2022}%
  \BibitemOpen
  \bibfield  {author} {\bibinfo {author} {\bibfnamefont {Y.}~\bibnamefont
  {Guan}}\ and\ \bibinfo {author} {\bibfnamefont {O.~V.}\ \bibnamefont
  {Yazyev}},\ }\bibfield  {title} {\bibinfo {title} {Electronic transport in
  graphene with out-of-plane disorder}\ }\href
  {https://doi.org/10.48550/arxiv.2210.16629} {10.48550/arxiv.2210.16629}
  (\bibinfo {year} {2022})\BibitemShut {NoStop}%
\bibitem [{\citenamefont {Pereira}\ and\ \citenamefont
  {Neto}(2009)}]{Pereira2009}%
  \BibitemOpen
  \bibfield  {author} {\bibinfo {author} {\bibfnamefont {V.~M.}\ \bibnamefont
  {Pereira}}\ and\ \bibinfo {author} {\bibfnamefont {A.~H.~C.}\ \bibnamefont
  {Neto}},\ }\bibfield  {title} {\bibinfo {title} {Strain engineering of
  graphene's electronic structure},\ }\href
  {https://doi.org/10.1103/physrevlett.103.046801} {\bibfield  {journal}
  {\bibinfo  {journal} {Physical Review Letters}\ }\textbf {\bibinfo {volume}
  {103}},\ \bibinfo {pages} {046801} (\bibinfo {year} {2009})}\BibitemShut
  {NoStop}%
\bibitem [{\citenamefont {Weiss}\ \emph {et~al.}(2012)\citenamefont {Weiss},
  \citenamefont {Zhou}, \citenamefont {Liao}, \citenamefont {Liu},
  \citenamefont {Jiang}, \citenamefont {Huang},\ and\ \citenamefont
  {Duan}}]{Weiss2012}%
  \BibitemOpen
  \bibfield  {author} {\bibinfo {author} {\bibfnamefont {N.~O.}\ \bibnamefont
  {Weiss}}, \bibinfo {author} {\bibfnamefont {H.}~\bibnamefont {Zhou}},
  \bibinfo {author} {\bibfnamefont {L.}~\bibnamefont {Liao}}, \bibinfo {author}
  {\bibfnamefont {Y.}~\bibnamefont {Liu}}, \bibinfo {author} {\bibfnamefont
  {S.}~\bibnamefont {Jiang}}, \bibinfo {author} {\bibfnamefont
  {Y.}~\bibnamefont {Huang}},\ and\ \bibinfo {author} {\bibfnamefont
  {X.}~\bibnamefont {Duan}},\ }\bibfield  {title} {\bibinfo {title} {Graphene:
  An emerging electronic material},\ }\href
  {https://doi.org/10.1002/adma.201201482} {\bibfield  {journal} {\bibinfo
  {journal} {Advanced Materials}\ }\textbf {\bibinfo {volume} {24}},\ \bibinfo
  {pages} {5782} (\bibinfo {year} {2012})}\BibitemShut {NoStop}%
\bibitem [{\citenamefont {Si}\ \emph {et~al.}(2016)\citenamefont {Si},
  \citenamefont {Sun},\ and\ \citenamefont {Liu}}]{Si2016}%
  \BibitemOpen
  \bibfield  {author} {\bibinfo {author} {\bibfnamefont {C.}~\bibnamefont
  {Si}}, \bibinfo {author} {\bibfnamefont {Z.}~\bibnamefont {Sun}},\ and\
  \bibinfo {author} {\bibfnamefont {F.}~\bibnamefont {Liu}},\ }\bibfield
  {title} {\bibinfo {title} {Strain engineering of graphene: a review},\ }\href
  {https://doi.org/10.1039/c5nr07755a} {\bibfield  {journal} {\bibinfo
  {journal} {Nanoscale}\ }\textbf {\bibinfo {volume} {8}},\ \bibinfo {pages}
  {3207} (\bibinfo {year} {2016})}\BibitemShut {NoStop}%
\bibitem [{\citenamefont {Neto}\ \emph {et~al.}(2009)\citenamefont {Neto},
  \citenamefont {Guinea}, \citenamefont {Peres}, \citenamefont {Novoselov},\
  and\ \citenamefont {Geim}}]{CastroNeto2009}%
  \BibitemOpen
  \bibfield  {author} {\bibinfo {author} {\bibfnamefont {A.~H.~C.}\
  \bibnamefont {Neto}}, \bibinfo {author} {\bibfnamefont {F.}~\bibnamefont
  {Guinea}}, \bibinfo {author} {\bibfnamefont {N.~M.~R.}\ \bibnamefont
  {Peres}}, \bibinfo {author} {\bibfnamefont {K.~S.}\ \bibnamefont
  {Novoselov}},\ and\ \bibinfo {author} {\bibfnamefont {A.~K.}\ \bibnamefont
  {Geim}},\ }\bibfield  {title} {\bibinfo {title} {The electronic properties of
  graphene},\ }\href {https://doi.org/10.1103/revmodphys.81.109} {\bibfield
  {journal} {\bibinfo  {journal} {Reviews of Modern Physics}\ }\textbf
  {\bibinfo {volume} {81}},\ \bibinfo {pages} {109} (\bibinfo {year}
  {2009})}\BibitemShut {NoStop}%
\bibitem [{\citenamefont {Suzuki}\ \emph {et~al.}(2017)\citenamefont {Suzuki},
  \citenamefont {Tanabe},\ and\ \citenamefont {Fujita}}]{Suzuki2017}%
  \BibitemOpen
  \bibfield  {author} {\bibinfo {author} {\bibfnamefont {A.}~\bibnamefont
  {Suzuki}}, \bibinfo {author} {\bibfnamefont {M.}~\bibnamefont {Tanabe}},\
  and\ \bibinfo {author} {\bibfnamefont {S.}~\bibnamefont {Fujita}},\
  }\bibfield  {title} {\bibinfo {title} {Electronic band structure of graphene
  based on the rectangular 4-atom unit cell},\ }\href
  {https://doi.org/10.4236/jmp.2017.84041} {\bibfield  {journal} {\bibinfo
  {journal} {Journal of Modern Physics}\ }\textbf {\bibinfo {volume} {08}},\
  \bibinfo {pages} {607} (\bibinfo {year} {2017})}\BibitemShut {NoStop}%
\bibitem [{\citenamefont {Georgoulea}\ \emph {et~al.}(2022)\citenamefont
  {Georgoulea}, \citenamefont {Power},\ and\ \citenamefont
  {Caffrey}}]{Georgoulea2022}%
  \BibitemOpen
  \bibfield  {author} {\bibinfo {author} {\bibfnamefont {N.~C.}\ \bibnamefont
  {Georgoulea}}, \bibinfo {author} {\bibfnamefont {S.~R.}\ \bibnamefont
  {Power}},\ and\ \bibinfo {author} {\bibfnamefont {N.~M.}\ \bibnamefont
  {Caffrey}},\ }\bibfield  {title} {\bibinfo {title} {Strain-induced stacking
  transition in bilayer graphene},\ }\href
  {https://doi.org/10.1088/1361-648x/ac965d} {\bibfield  {journal} {\bibinfo
  {journal} {Journal of Physics: Condensed Matter}\ }\textbf {\bibinfo {volume}
  {34}},\ \bibinfo {pages} {475302} (\bibinfo {year} {2022})}\BibitemShut
  {NoStop}%
\bibitem [{\citenamefont {Choi}\ \emph {et~al.}(2010)\citenamefont {Choi},
  \citenamefont {Jhi},\ and\ \citenamefont {Son}}]{Choi2010}%
  \BibitemOpen
  \bibfield  {author} {\bibinfo {author} {\bibfnamefont {S.-M.}\ \bibnamefont
  {Choi}}, \bibinfo {author} {\bibfnamefont {S.-H.}\ \bibnamefont {Jhi}},\ and\
  \bibinfo {author} {\bibfnamefont {Y.-W.}\ \bibnamefont {Son}},\ }\bibfield
  {title} {\bibinfo {title} {Controlling energy gap of bilayer graphene by
  strain},\ }\href {https://doi.org/10.1021/nl101617x} {\bibfield  {journal}
  {\bibinfo  {journal} {Nano Letters}\ }\textbf {\bibinfo {volume} {10}},\
  \bibinfo {pages} {3486} (\bibinfo {year} {2010})}\BibitemShut {NoStop}%
\bibitem [{\citenamefont {Kennes}\ \emph {et~al.}(2020)\citenamefont {Kennes},
  \citenamefont {Xian}, \citenamefont {Claassen},\ and\ \citenamefont
  {Rubio}}]{Kennes2020}%
  \BibitemOpen
  \bibfield  {author} {\bibinfo {author} {\bibfnamefont {D.~M.}\ \bibnamefont
  {Kennes}}, \bibinfo {author} {\bibfnamefont {L.}~\bibnamefont {Xian}},
  \bibinfo {author} {\bibfnamefont {M.}~\bibnamefont {Claassen}},\ and\
  \bibinfo {author} {\bibfnamefont {A.}~\bibnamefont {Rubio}},\ }\bibfield
  {title} {\bibinfo {title} {One-dimensional flat bands in twisted bilayer
  germanium selenide},\ }\href {https://doi.org/10.1038/s41467-020-14947-0}
  {\bibfield  {journal} {\bibinfo  {journal} {Nature Communications}\ }\textbf
  {\bibinfo {volume} {11}},\ \bibinfo {pages} {1124} (\bibinfo {year}
  {2020})}\BibitemShut {NoStop}%
\bibitem [{\citenamefont {Bonfanti}\ \emph {et~al.}(2018)\citenamefont
  {Bonfanti}, \citenamefont {Achilli},\ and\ \citenamefont
  {Martinazzo}}]{Bonfanti2018}%
  \BibitemOpen
  \bibfield  {author} {\bibinfo {author} {\bibfnamefont {M.}~\bibnamefont
  {Bonfanti}}, \bibinfo {author} {\bibfnamefont {S.}~\bibnamefont {Achilli}},\
  and\ \bibinfo {author} {\bibfnamefont {R.}~\bibnamefont {Martinazzo}},\
  }\href {https://doi.org/10.1088/1361-648X/aac89f} {\bibfield  {journal}
  {\bibinfo  {journal} {Journal of Physics: Condensed Matter}\ }\textbf
  {\bibinfo {volume} {30}},\ \bibinfo {pages} {283002} (\bibinfo {year}
  {2018})}\BibitemShut {NoStop}%
\bibitem [{\citenamefont {Park}\ \emph {et~al.}(2003)\citenamefont {Park},
  \citenamefont {Srivastava},\ and\ \citenamefont {Cho}}]{Park2003}%
  \BibitemOpen
  \bibfield  {author} {\bibinfo {author} {\bibfnamefont {S.}~\bibnamefont
  {Park}}, \bibinfo {author} {\bibfnamefont {D.}~\bibnamefont {Srivastava}},\
  and\ \bibinfo {author} {\bibfnamefont {K.}~\bibnamefont {Cho}},\ }\bibfield
  {title} {\bibinfo {title} {Generalized chemical reactivity of curved
  surfaces: Carbon nanotubes},\ }\href {https://doi.org/10.1021/nl0342747}
  {\bibfield  {journal} {\bibinfo  {journal} {Nano Letters}\ }\textbf {\bibinfo
  {volume} {3}},\ \bibinfo {pages} {1273} (\bibinfo {year} {2003})}\BibitemShut
  {NoStop}%
\bibitem [{\citenamefont {Balog}\ \emph {et~al.}(2009)\citenamefont {Balog},
  \citenamefont {Jørgensen}, \citenamefont {Wells}, \citenamefont
  {Lægsgaard}, \citenamefont {Hofmann}, \citenamefont {Besenbacher},\ and\
  \citenamefont {Hornekær}}]{Balog2009}%
  \BibitemOpen
  \bibfield  {author} {\bibinfo {author} {\bibfnamefont {R.}~\bibnamefont
  {Balog}}, \bibinfo {author} {\bibfnamefont {B.}~\bibnamefont {Jørgensen}},
  \bibinfo {author} {\bibfnamefont {J.}~\bibnamefont {Wells}}, \bibinfo
  {author} {\bibfnamefont {E.}~\bibnamefont {Lægsgaard}}, \bibinfo {author}
  {\bibfnamefont {P.}~\bibnamefont {Hofmann}}, \bibinfo {author} {\bibfnamefont
  {F.}~\bibnamefont {Besenbacher}},\ and\ \bibinfo {author} {\bibfnamefont
  {L.}~\bibnamefont {Hornekær}},\ }\bibfield  {title} {\bibinfo {title}
  {Atomic hydrogen adsorbate structures on graphene},\ }\href
  {https://doi.org/10.1021/ja902714h} {\bibfield  {journal} {\bibinfo
  {journal} {Journal of the American Chemical Society}\ }\textbf {\bibinfo
  {volume} {131}},\ \bibinfo {pages} {8744} (\bibinfo {year}
  {2009})}\BibitemShut {NoStop}%
\bibitem [{\citenamefont {Balog}\ \emph {et~al.}(2010)\citenamefont {Balog},
  \citenamefont {J{\o}rgensen}, \citenamefont {Nilsson}, \citenamefont
  {Andersen}, \citenamefont {Rienks}, \citenamefont {Bianchi}, \citenamefont
  {Fanetti}, \citenamefont {L{\ae}gsgaard}, \citenamefont {Baraldi},
  \citenamefont {Lizzit}, \citenamefont {Sljivancanin}, \citenamefont
  {Besenbacher}, \citenamefont {Hammer}, \citenamefont {Pedersen},
  \citenamefont {Hofmann},\ and\ \citenamefont {Hornek{\ae}r}}]{Balog2010}%
  \BibitemOpen
  \bibfield  {author} {\bibinfo {author} {\bibfnamefont {R.}~\bibnamefont
  {Balog}}, \bibinfo {author} {\bibfnamefont {B.}~\bibnamefont {J{\o}rgensen}},
  \bibinfo {author} {\bibfnamefont {L.}~\bibnamefont {Nilsson}}, \bibinfo
  {author} {\bibfnamefont {M.}~\bibnamefont {Andersen}}, \bibinfo {author}
  {\bibfnamefont {E.}~\bibnamefont {Rienks}}, \bibinfo {author} {\bibfnamefont
  {M.}~\bibnamefont {Bianchi}}, \bibinfo {author} {\bibfnamefont
  {M.}~\bibnamefont {Fanetti}}, \bibinfo {author} {\bibfnamefont
  {E.}~\bibnamefont {L{\ae}gsgaard}}, \bibinfo {author} {\bibfnamefont
  {A.}~\bibnamefont {Baraldi}}, \bibinfo {author} {\bibfnamefont
  {S.}~\bibnamefont {Lizzit}}, \bibinfo {author} {\bibfnamefont
  {Z.}~\bibnamefont {Sljivancanin}}, \bibinfo {author} {\bibfnamefont
  {F.}~\bibnamefont {Besenbacher}}, \bibinfo {author} {\bibfnamefont
  {B.}~\bibnamefont {Hammer}}, \bibinfo {author} {\bibfnamefont {T.~G.}\
  \bibnamefont {Pedersen}}, \bibinfo {author} {\bibfnamefont {P.}~\bibnamefont
  {Hofmann}},\ and\ \bibinfo {author} {\bibfnamefont {L.}~\bibnamefont
  {Hornek{\ae}r}},\ }\bibfield  {title} {\bibinfo {title} {Bandgap opening in
  graphene induced by patterned hydrogen adsorption},\ }\href
  {https://doi.org/10.1038/nmat2710} {\bibfield  {journal} {\bibinfo  {journal}
  {Nature Materials}\ }\textbf {\bibinfo {volume} {9}},\ \bibinfo {pages} {315}
  (\bibinfo {year} {2010})}\BibitemShut {NoStop}%
\bibitem [{\citenamefont {Elias}\ \emph {et~al.}(2009)\citenamefont {Elias},
  \citenamefont {Nair}, \citenamefont {Mohiuddin}, \citenamefont {Morozov},
  \citenamefont {Blake}, \citenamefont {Halsall}, \citenamefont {Ferrari},
  \citenamefont {Boukhvalov}, \citenamefont {Katsnelson}, \citenamefont
  {Geim},\ and\ \citenamefont {Novoselov}}]{Elias2009}%
  \BibitemOpen
  \bibfield  {author} {\bibinfo {author} {\bibfnamefont {D.~C.}\ \bibnamefont
  {Elias}}, \bibinfo {author} {\bibfnamefont {R.~R.}\ \bibnamefont {Nair}},
  \bibinfo {author} {\bibfnamefont {T.~M.~G.}\ \bibnamefont {Mohiuddin}},
  \bibinfo {author} {\bibfnamefont {S.~V.}\ \bibnamefont {Morozov}}, \bibinfo
  {author} {\bibfnamefont {P.}~\bibnamefont {Blake}}, \bibinfo {author}
  {\bibfnamefont {M.~P.}\ \bibnamefont {Halsall}}, \bibinfo {author}
  {\bibfnamefont {A.~C.}\ \bibnamefont {Ferrari}}, \bibinfo {author}
  {\bibfnamefont {D.~W.}\ \bibnamefont {Boukhvalov}}, \bibinfo {author}
  {\bibfnamefont {M.~I.}\ \bibnamefont {Katsnelson}}, \bibinfo {author}
  {\bibfnamefont {A.~K.}\ \bibnamefont {Geim}},\ and\ \bibinfo {author}
  {\bibfnamefont {K.~S.}\ \bibnamefont {Novoselov}},\ }\bibfield  {title}
  {\bibinfo {title} {Control of graphene's properties by reversible
  hydrogenation: Evidence for graphane},\ }\href
  {https://doi.org/10.1126/science.1167130} {\bibfield  {journal} {\bibinfo
  {journal} {Science}\ }\textbf {\bibinfo {volume} {323}},\ \bibinfo {pages}
  {610} (\bibinfo {year} {2009})}\BibitemShut {NoStop}%
\bibitem [{\citenamefont {Lieb}(1989)}]{Lieb1989}%
  \BibitemOpen
  \bibfield  {author} {\bibinfo {author} {\bibfnamefont {E.~H.}\ \bibnamefont
  {Lieb}},\ }\bibfield  {title} {\bibinfo {title} {Two theorems on the hubbard
  model},\ }\href {https://doi.org/10.1103/PhysRevLett.62.1201} {\bibfield
  {journal} {\bibinfo  {journal} {Physical Review Letters}\ }\textbf {\bibinfo
  {volume} {62}},\ \bibinfo {pages} {1201} (\bibinfo {year}
  {1989})}\BibitemShut {NoStop}%
\bibitem [{\citenamefont {Casolo}\ \emph {et~al.}(2009)\citenamefont {Casolo},
  \citenamefont {Løvvik}, \citenamefont {Martinazzo},\ and\ \citenamefont
  {Tantardini}}]{Casolo2009}%
  \BibitemOpen
  \bibfield  {author} {\bibinfo {author} {\bibfnamefont {S.}~\bibnamefont
  {Casolo}}, \bibinfo {author} {\bibfnamefont {O.~M.}\ \bibnamefont {Løvvik}},
  \bibinfo {author} {\bibfnamefont {R.}~\bibnamefont {Martinazzo}},\ and\
  \bibinfo {author} {\bibfnamefont {G.~F.}\ \bibnamefont {Tantardini}},\
  }\bibfield  {title} {\bibinfo {title} {Understanding adsorption of hydrogen
  atoms on graphene},\ }\href {https://doi.org/10.1063/1.3072333} {\bibfield
  {journal} {\bibinfo  {journal} {The Journal of Chemical Physics}\ }\textbf
  {\bibinfo {volume} {130}},\ \bibinfo {pages} {054704} (\bibinfo {year}
  {2009})}\BibitemShut {NoStop}%
\bibitem [{\citenamefont {Yazyev}\ and\ \citenamefont
  {Helm}(2007)}]{Yazyev2007}%
  \BibitemOpen
  \bibfield  {author} {\bibinfo {author} {\bibfnamefont {O.~V.}\ \bibnamefont
  {Yazyev}}\ and\ \bibinfo {author} {\bibfnamefont {L.}~\bibnamefont {Helm}},\
  }\bibfield  {title} {\bibinfo {title} {Defect-induced magnetism in
  graphene},\ }\href {https://doi.org/10.1103/PhysRevB.75.125408} {\bibfield
  {journal} {\bibinfo  {journal} {Physical Review B}\ }\textbf {\bibinfo
  {volume} {75}},\ \bibinfo {pages} {125408} (\bibinfo {year}
  {2007})}\BibitemShut {NoStop}%
\bibitem [{\citenamefont {Pizzochero}\ and\ \citenamefont
  {Kaxiras}(2022)}]{Pizzochero2022}%
  \BibitemOpen
  \bibfield  {author} {\bibinfo {author} {\bibfnamefont {M.}~\bibnamefont
  {Pizzochero}}\ and\ \bibinfo {author} {\bibfnamefont {E.}~\bibnamefont
  {Kaxiras}},\ }\bibfield  {title} {\bibinfo {title} {Hydrogen atoms on zigzag
  graphene nanoribbons: Chemistry and magnetism meet at the edge},\ }\href
  {https://doi.org/10.1021/acs.nanolett.1c04362} {\bibfield  {journal}
  {\bibinfo  {journal} {Nano Letters}\ }\textbf {\bibinfo {volume} {22}},\
  \bibinfo {pages} {1922} (\bibinfo {year} {2022})}\BibitemShut {NoStop}%
\bibitem [{\citenamefont {González-Herrero}\ \emph {et~al.}(2016)\citenamefont
  {González-Herrero}, \citenamefont {Gómez-Rodríguez}, \citenamefont
  {Mallet}, \citenamefont {Moaied}, \citenamefont {Palacios}, \citenamefont
  {Salgado}, \citenamefont {Ugeda}, \citenamefont {Veuillen}, \citenamefont
  {Yndurain},\ and\ \citenamefont {Brihuega}}]{Gonzalez-Herrero2016}%
  \BibitemOpen
  \bibfield  {author} {\bibinfo {author} {\bibfnamefont {H.}~\bibnamefont
  {González-Herrero}}, \bibinfo {author} {\bibfnamefont {J.~M.}\ \bibnamefont
  {Gómez-Rodríguez}}, \bibinfo {author} {\bibfnamefont {P.}~\bibnamefont
  {Mallet}}, \bibinfo {author} {\bibfnamefont {M.}~\bibnamefont {Moaied}},
  \bibinfo {author} {\bibfnamefont {J.~J.}\ \bibnamefont {Palacios}}, \bibinfo
  {author} {\bibfnamefont {C.}~\bibnamefont {Salgado}}, \bibinfo {author}
  {\bibfnamefont {M.~M.}\ \bibnamefont {Ugeda}}, \bibinfo {author}
  {\bibfnamefont {J.-Y.}\ \bibnamefont {Veuillen}}, \bibinfo {author}
  {\bibfnamefont {F.}~\bibnamefont {Yndurain}},\ and\ \bibinfo {author}
  {\bibfnamefont {I.}~\bibnamefont {Brihuega}},\ }\bibfield  {title} {\bibinfo
  {title} {Atomic-scale control of graphene magnetism by using hydrogen
  atoms},\ }\href {https://doi.org/10.1126/science.aad8038} {\bibfield
  {journal} {\bibinfo  {journal} {Science}\ }\textbf {\bibinfo {volume}
  {352}},\ \bibinfo {pages} {437} (\bibinfo {year} {2016})}\BibitemShut
  {NoStop}%
\bibitem [{\citenamefont {Nguyen}\ \emph {et~al.}(2022)\citenamefont {Nguyen},
  \citenamefont {Letartre}, \citenamefont {Drouard}, \citenamefont
  {Viktorovitch}, \citenamefont {Nguyen},\ and\ \citenamefont
  {Nguyen}}]{Nguyen2022}%
  \BibitemOpen
  \bibfield  {author} {\bibinfo {author} {\bibfnamefont {D.~X.}\ \bibnamefont
  {Nguyen}}, \bibinfo {author} {\bibfnamefont {X.}~\bibnamefont {Letartre}},
  \bibinfo {author} {\bibfnamefont {E.}~\bibnamefont {Drouard}}, \bibinfo
  {author} {\bibfnamefont {P.}~\bibnamefont {Viktorovitch}}, \bibinfo {author}
  {\bibfnamefont {H.~C.}\ \bibnamefont {Nguyen}},\ and\ \bibinfo {author}
  {\bibfnamefont {H.~S.}\ \bibnamefont {Nguyen}},\ }\bibfield  {title}
  {\bibinfo {title} {Magic configurations in moir{\'{e}} superlattice of
  bilayer photonic crystals: Almost-perfect flatbands and unconventional
  localization},\ }\href {https://doi.org/10.1103/physrevresearch.4.l032031}
  {\bibfield  {journal} {\bibinfo  {journal} {Physical Review Research}\
  }\textbf {\bibinfo {volume} {4}},\ \bibinfo {pages} {L032031} (\bibinfo
  {year} {2022})}\BibitemShut {NoStop}%
\bibitem [{\citenamefont {Hsieh}\ \emph {et~al.}(2023)\citenamefont {Hsieh},
  \citenamefont {Halbertal}, \citenamefont {Finney}, \citenamefont {Zhu},
  \citenamefont {Gerber}, \citenamefont {Pizzochero}, \citenamefont
  {Kucukbenli}, \citenamefont {Schleder}, \citenamefont {Angeli}, \citenamefont
  {Watanabe}, \citenamefont {Taniguchi}, \citenamefont {Kim}, \citenamefont
  {Kaxiras}, \citenamefont {Hone}, \citenamefont {Dean},\ and\ \citenamefont
  {Basov}}]{Hsieh2022}%
  \BibitemOpen
  \bibfield  {author} {\bibinfo {author} {\bibfnamefont {V.}~\bibnamefont
  {Hsieh}}, \bibinfo {author} {\bibfnamefont {D.}~\bibnamefont {Halbertal}},
  \bibinfo {author} {\bibfnamefont {N.~R.}\ \bibnamefont {Finney}}, \bibinfo
  {author} {\bibfnamefont {Z.}~\bibnamefont {Zhu}}, \bibinfo {author}
  {\bibfnamefont {E.}~\bibnamefont {Gerber}}, \bibinfo {author} {\bibfnamefont
  {M.}~\bibnamefont {Pizzochero}}, \bibinfo {author} {\bibfnamefont
  {E.}~\bibnamefont {Kucukbenli}}, \bibinfo {author} {\bibfnamefont {G.~R.}\
  \bibnamefont {Schleder}}, \bibinfo {author} {\bibfnamefont {M.}~\bibnamefont
  {Angeli}}, \bibinfo {author} {\bibfnamefont {K.}~\bibnamefont {Watanabe}},
  \bibinfo {author} {\bibfnamefont {T.}~\bibnamefont {Taniguchi}}, \bibinfo
  {author} {\bibfnamefont {E.-A.}\ \bibnamefont {Kim}}, \bibinfo {author}
  {\bibfnamefont {E.}~\bibnamefont {Kaxiras}}, \bibinfo {author} {\bibfnamefont
  {J.}~\bibnamefont {Hone}}, \bibinfo {author} {\bibfnamefont {C.~R.}\
  \bibnamefont {Dean}},\ and\ \bibinfo {author} {\bibfnamefont {D.~N.}\
  \bibnamefont {Basov}},\ }\bibfield  {title} {\bibinfo {title}
  {Domain-dependent surface adhesion in twisted few-layer graphene: Platform
  for moir{\'{e}}-assisted chemistry},\ }\href
  {https://doi.org/10.1021/acs.nanolett.2c04137} {\bibfield  {journal}
  {\bibinfo  {journal} {Nano Letters}\ }\textbf {\bibinfo {volume} {23}},\
  \bibinfo {pages} {3137} (\bibinfo {year} {2023})}\BibitemShut {NoStop}%
\end{thebibliography}
%

\clearpage


\section*{Supplementary material}

# One-Dimensional Moiré Physics and Chemistry in Heterostrained Bilayer Graphene

Gabriel R. Schleder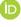,[1, *] Michele Pizzochero,[1, *] and Efthimios Kaxiras[1, 2]

[1] *John A. Paulson School of Engineering and Applied Sciences,*
*Harvard University, Cambridge, MA 02138, United States*
[2] *Department of Physics, Harvard University, Cambridge, MA 02138, United States*

## SUPPORTING NOTE 1

### Atomistic relaxation calculations

The forcefield calculations are performed using the LAMMPS software [1]. For intralayer interactions, we use the second-generation reactive empirical bond-order (REBO) [2] potential. For interlayer interactions, we use a Kolmogorov–Crespi (KC) registry-dependent interlayer potential for graphitic systems [3] that has been re-parametrized to reproduce several observables calculated within DFT+vdW(rVV10) [4] as a function of interlayer distance and bilayer relative shift. The minimizations are performed using the *FIRE* method, with a force tolerance of $10^{-4}$ eV/Å, using the DFT lattice constants. The KC potential file and LAMMPS example input files for relaxation are available with the *Twister* package [5].

## SUPPORTING NOTE 2

### Band structure calculations details

We perform calculations based on density-functional theory (DFT), as implemented in the SIESTA software [6]. We use the generalized gradient approximation (GGA) to the exchange-correlation functional as devised by Perdew, Burke, and Ernzerhof (PBE) [7]. We use double-$\zeta$ polarized (DZP) numerical atomic orbitals (NAOs) basis sets for the expansion of the wavefunctions, a cut-off of 300 Ry for the real-space grid, and a threshold of $5 \times 10^{-4}$ eV for the Hamiltonian and density matrix self-consistent convergence. We use norm-conserving pseudopotentials [8] in the PSML format [9] obtained from *pseudo-dojo* [10]. For generating the charge density to be used in the band structure calculations, we sample the Brillouin zone with a mesh of $M \times 7 \times 1$ $k$-points, where $M$ maintains a density larger than $4.5/\text{Å}^{-1}$.

**Folding effect on unstrained bilayer graphene**

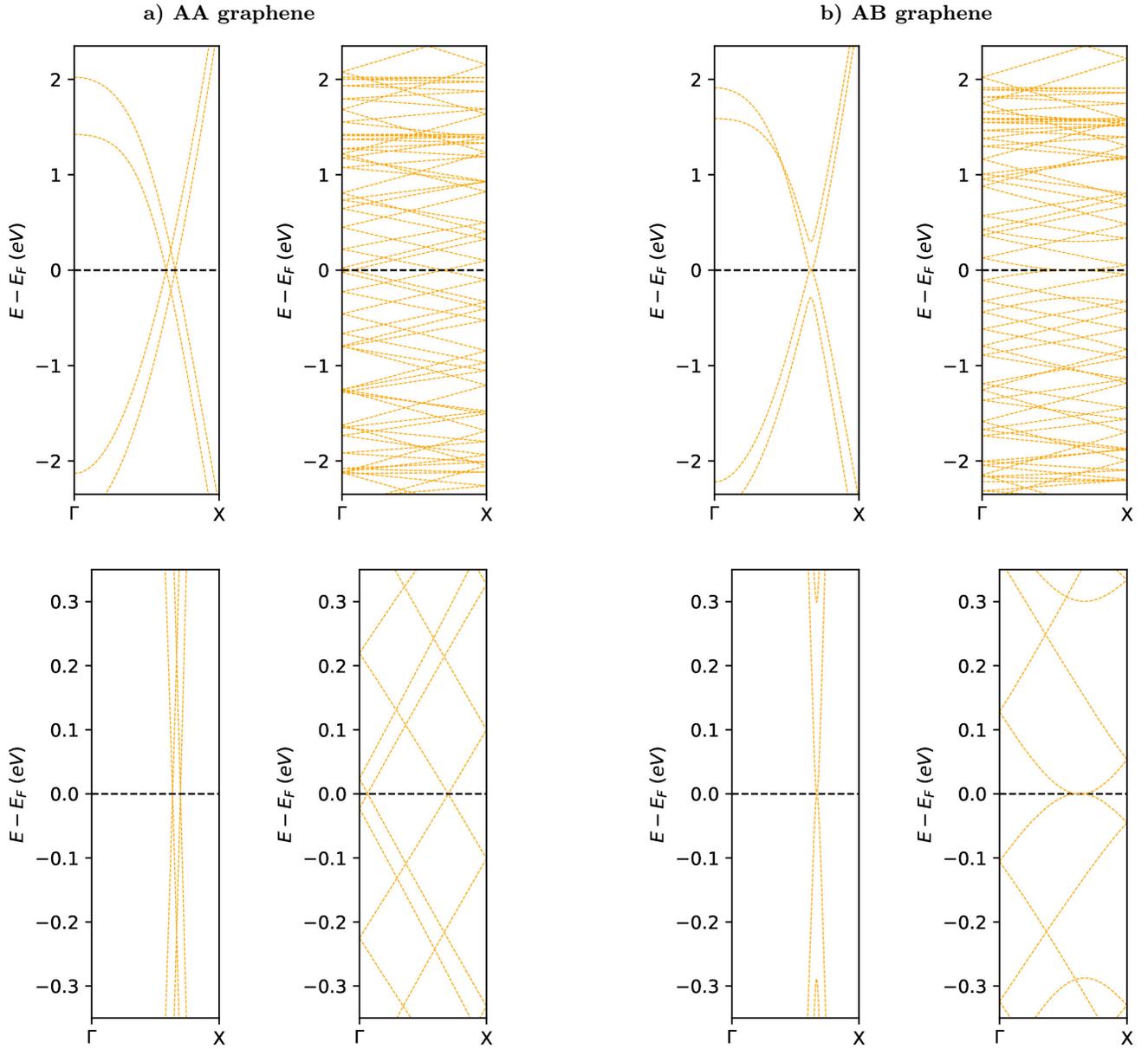

FIG. 1. Band folding of a) AA and b) AB graphene in the rectangular cell. Each pair of figures compare the 1×1 cell (left) with the 20×1 supercell (right). The bottom pair of figures shows a narrower energy range than the corresponding figure above it.

## Band structures of armchair-heterostrained bilayer graphene

The band structures of armchair-heterostrained bilayer graphene are calculated with the Moon–Koshino Slater–Koster type tight-binding model [11, 12], using the *letb* python implementation [13].

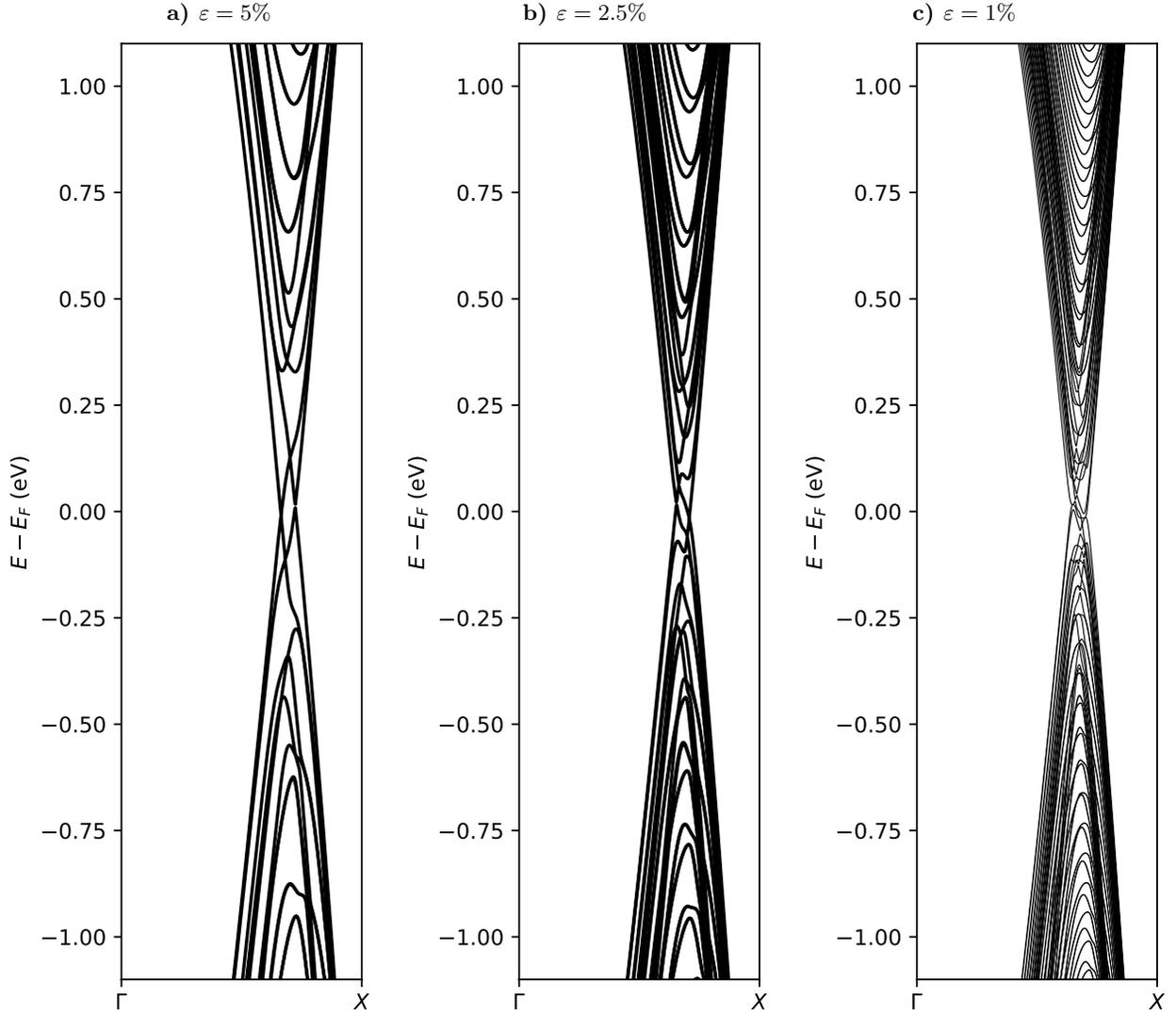

**a)** $\varepsilon = 5\%$          **b)** $\varepsilon = 2.5\%$          **c)** $\varepsilon = 1\%$

FIG. 2. Band structures of armchair-heterostrained bilayer graphene with different strain magnitudes: a) $\varepsilon = 5\%$, b) $\varepsilon = 2.5\%$, and c) $\varepsilon = 1\%$. All structures are metallic with different contributions of bilayer graphene local stackings.

**Band structures and properties of rigid heterostrained systems**

In Fig. 3, we show the band structure for the rigidly strained systems, where the strain is uniformly distributed along the strained direction. The magnitude of the strain ranges from $\varepsilon = -5\%$ (compressive) to $\varepsilon = +5\%$ (tensile). For both compressive and tensile strain, two sets of isolated valence and conduction bands form, having both a gap between them ($\Delta_{vc}$) and a gap to the band below or above it ($\Delta_{v-1}$ or $\Delta_{c+1}$). The width of these moiré bands ($w_v$ and $w_c$) decreases with the magnitude of the strain, being virtually zero when strains are about 1% while maintaining sizable gaps.

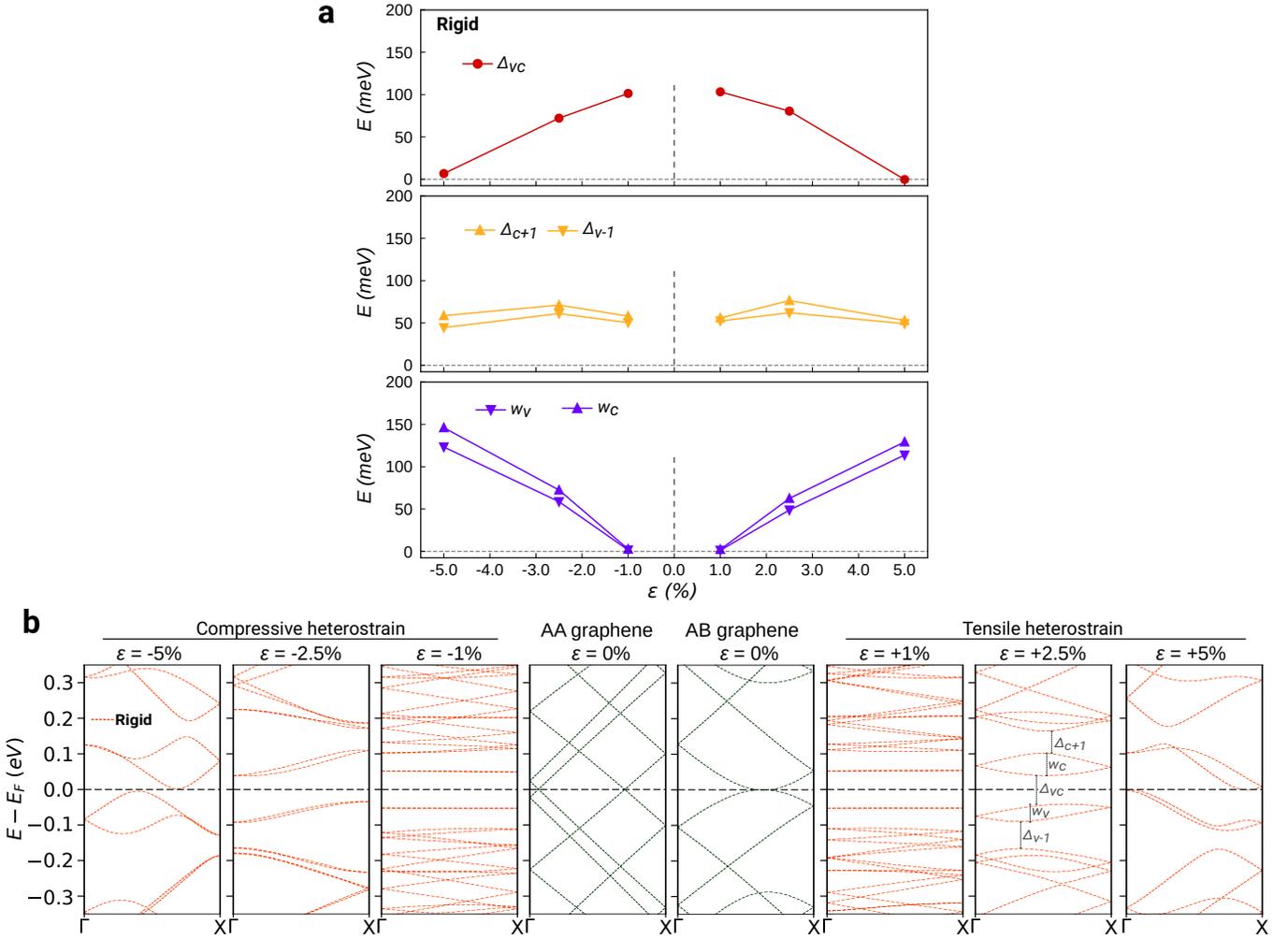

FIG. 3. Electronic structure of rigid hBLG systems. **a)** Summary of electronic properties of the relaxed systems: valence–conduction band gap $\Delta_{vc}$, gap to the band below or above it $\Delta_{v-1}$ or $\Delta_{c+1}$, and the width of the valence and conduction moiré bands $w_v$ and $w_c$. **b)** Band structures of rigid hBLGs, i.e., without relaxation effects. The results contrast with the relaxed ones presented in the main text.

# SUPPORTING NOTE 3

## Moiré chemistry of heterostrain bilayer graphene

To gain insight into the chemistry of heterostrain bilayer graphene, we perform calculations based on spin-polarized density-functional theory (DFT), as implemented in the Vienna *Ab Initio* Simulation Package (VASP) [14–16]. We resort to the generalized gradient approximation (GGA) to the exchange-correlation functional devised by Perdew, Burke, and Ernzerhof (PBE) [7]. For this set of calculations, we rely on the model of HBG shown in Figure 1a in the main text. This model is subjected to a compressive heterostrain of 5% and contains 492 carbon atoms. A vacuum layer of 20 Å is introduced in the out-of-plane direction to avoid artificial interactions between periodic replicas. Electron-core interactions are described through the projector augmented wave method (PAW) [17]. The Kohn-Sham wavefunctions of the valence electrons are expanded in a plane-wave basis set with a kinetic energy cutoff of 400 eV. The use of a plane-wave basis set (instead of a local basis set, cf. Supporting Note 2) prevents the basis-set superposition error (BSSE), which was shown to yield an incorrect description of the energetics of the hydrogen adsorption on graphene [18]. We sample the Brillouin zone with a mesh of $1 \times 3 \times 1$ $k$-points. The atomic coordinates are optimized until the maximum force component is lower than 0.04 eV/Å.

## SUPPORTING NOTE 4

**Atomistic relaxation: structures before and after relaxation**

**Zigzag heterostrained moiré structures: periodic simulation cells**

$\varepsilon = -0.05$: rigid moiré, i.e., unrelaxed. Top and both side views.

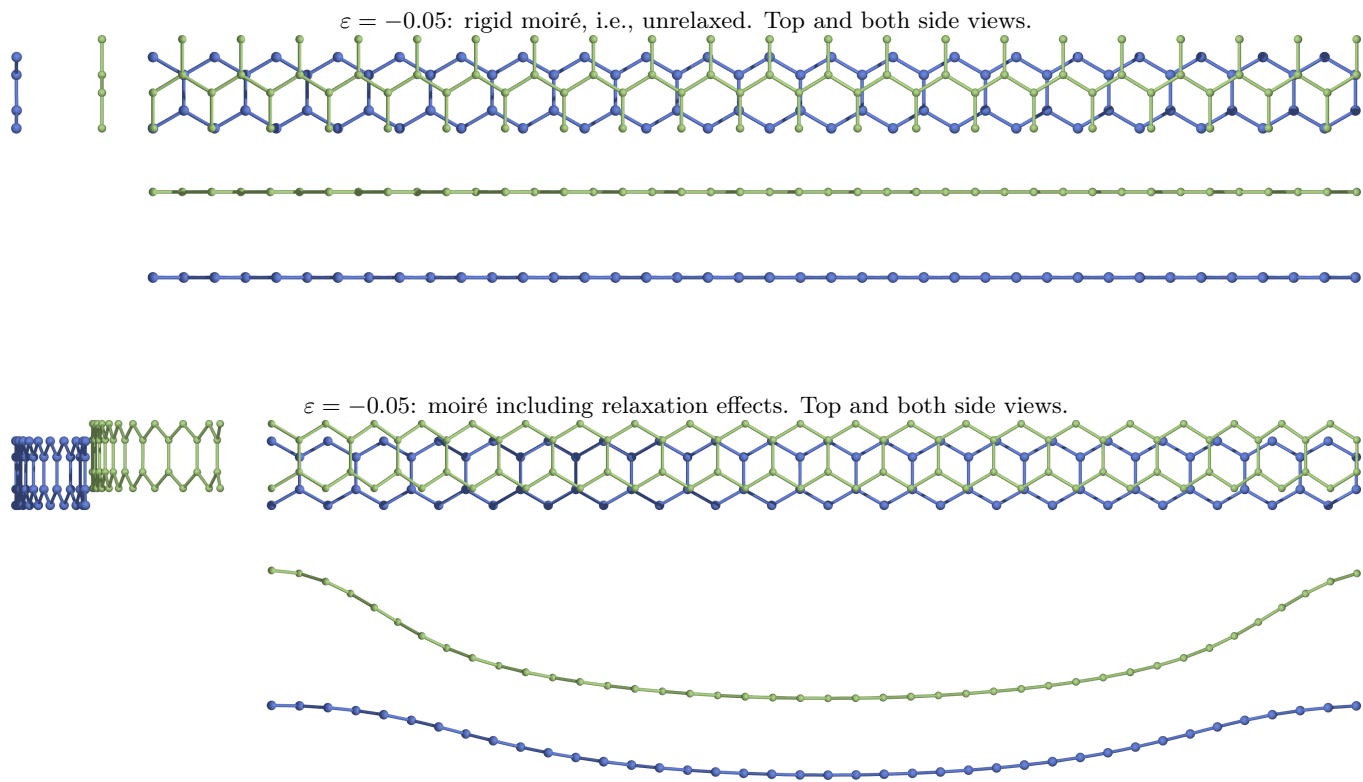

$\varepsilon = -0.05$: moiré including relaxation effects. Top and both side views.

FIG. 4. Structures of rigid and relaxed hBLGs, i.e., without and with relaxation effects.

$\varepsilon = -0.0333$: rigid moiré, i.e., unrelaxed. Top and both side views.

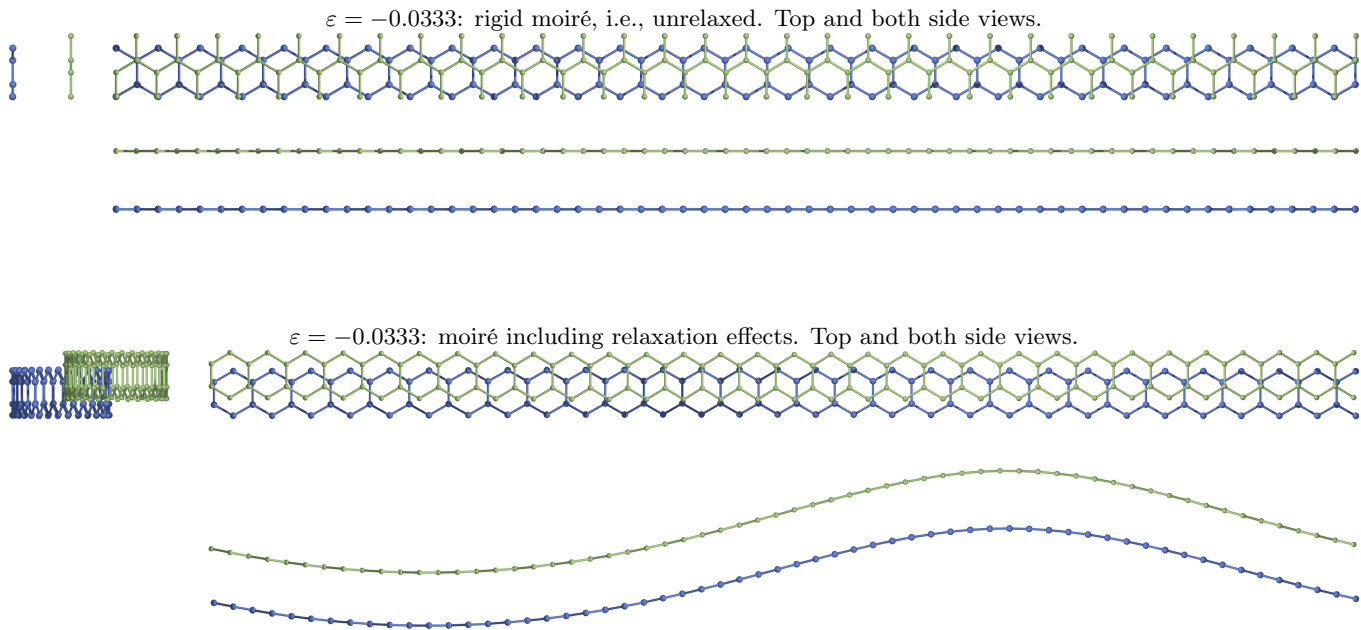

$\varepsilon = -0.0333$: moiré including relaxation effects. Top and both side views.

FIG. 5. Structures of rigid and relaxed hBLGs, i.e., without and with relaxation effects.

$\varepsilon = -0.025$: rigid moiré, i.e., unrelaxed. Top and both side views.

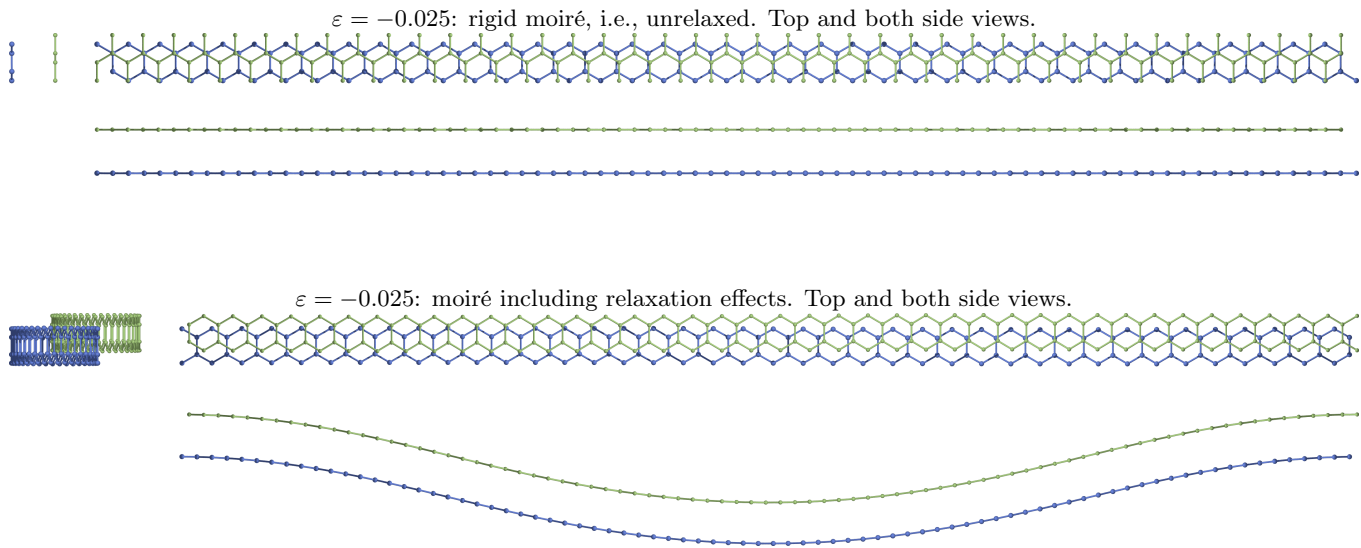

$\varepsilon = -0.025$: moiré including relaxation effects. Top and both side views.

FIG. 6. Structures of rigid and relaxed hBLGs, i.e., without and with relaxation effects.

$\varepsilon = -0.01$: rigid moiré, i.e., unrelaxed. Top and both side views.

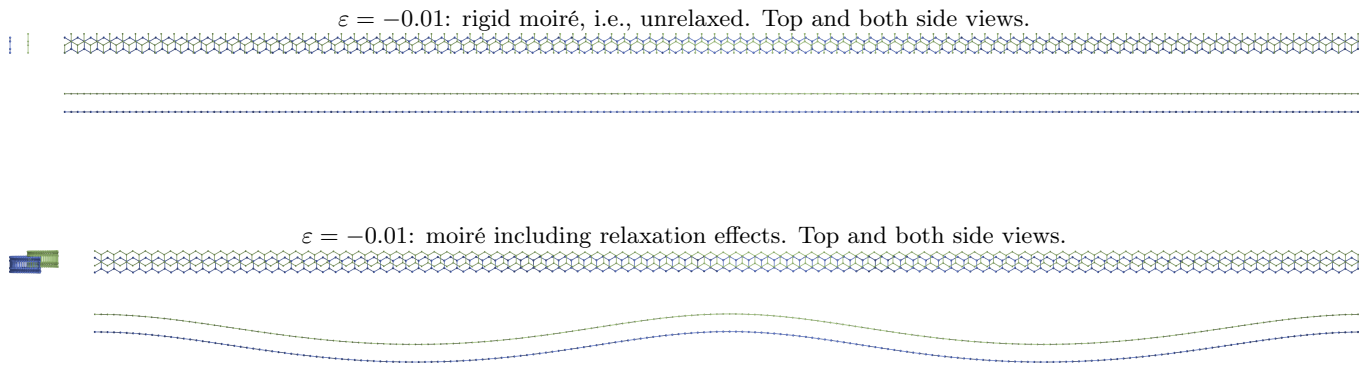

$\varepsilon = -0.01$: moiré including relaxation effects. Top and both side views.

FIG. 7. Structures of rigid and relaxed hBLGs, i.e., without and with relaxation effects.

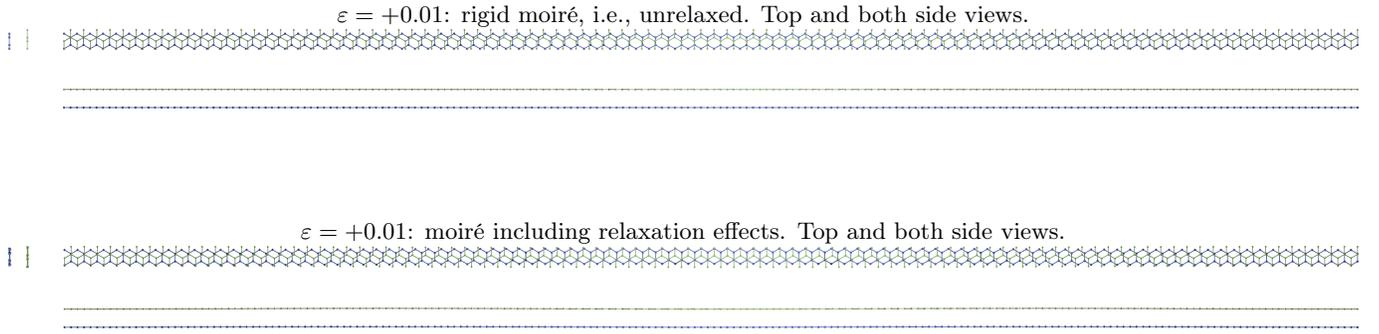

FIG. 8. Structures of rigid and relaxed hBLGs, i.e., without and with relaxation effects.

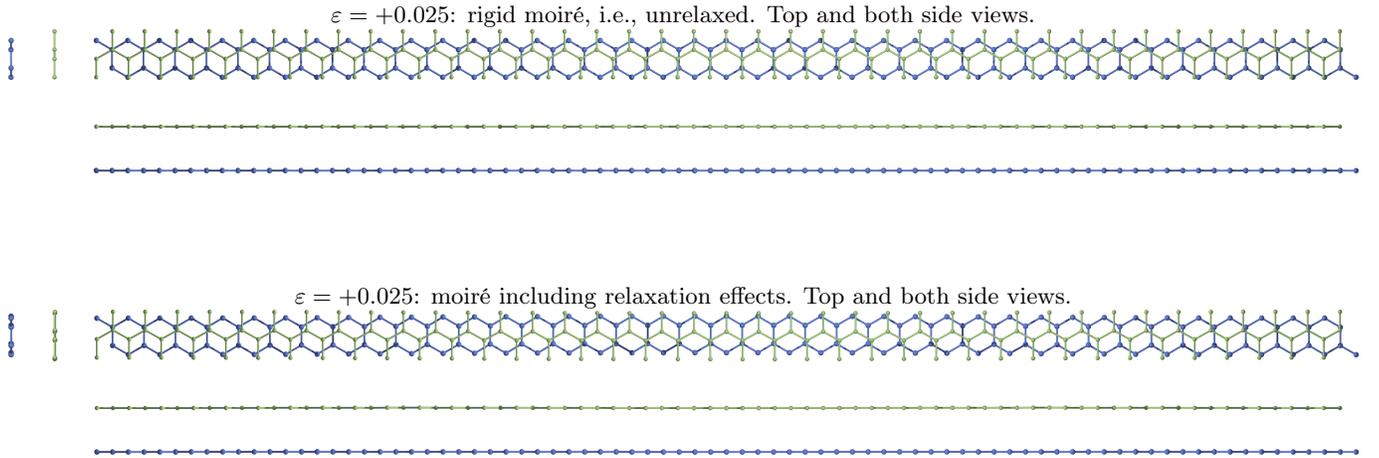

FIG. 9. Structures of rigid and relaxed hBLGs, i.e., without and with relaxation effects.

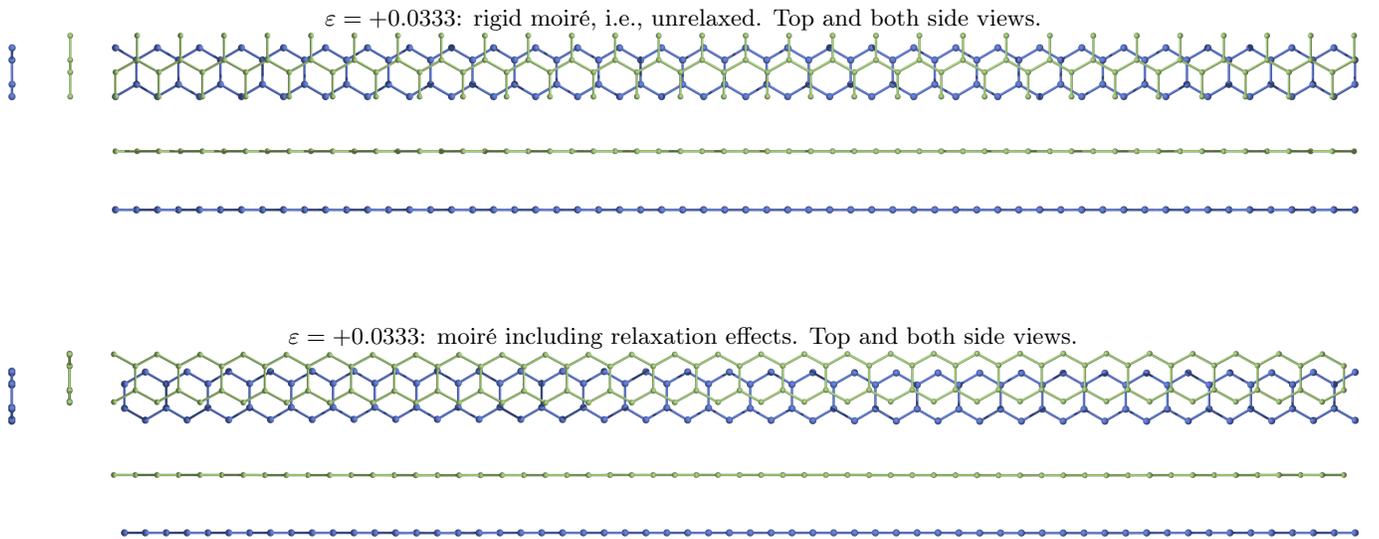

FIG. 10. Structures of rigid and relaxed hBLGs, i.e., without and with relaxation effects.

$\varepsilon = +0.05$: rigid moiré, i.e., unrelaxed. Top and both side views.

$\varepsilon = +0.05$: moiré including relaxation effects. Top and both side views.

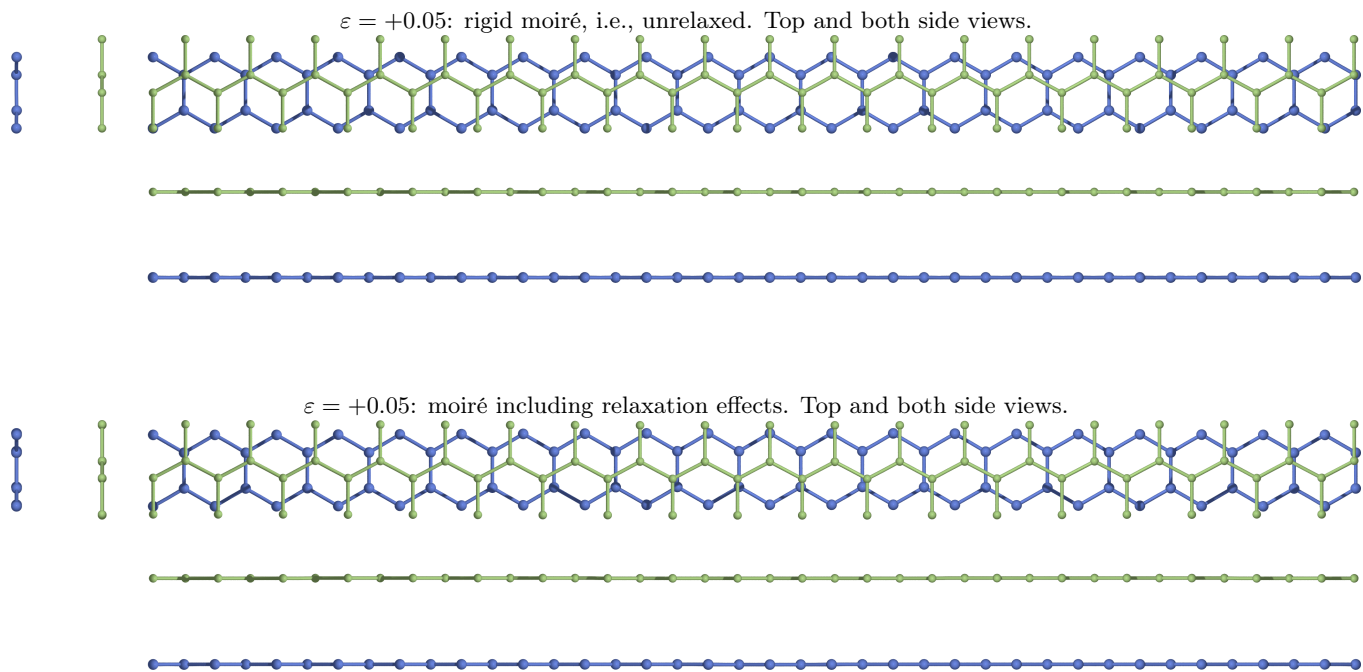

FIG. 11. Structures of rigid and relaxed hBLGs, i.e., without and with relaxation effects.

**Armchair heterostrained moiré structures: periodic simulation cells**

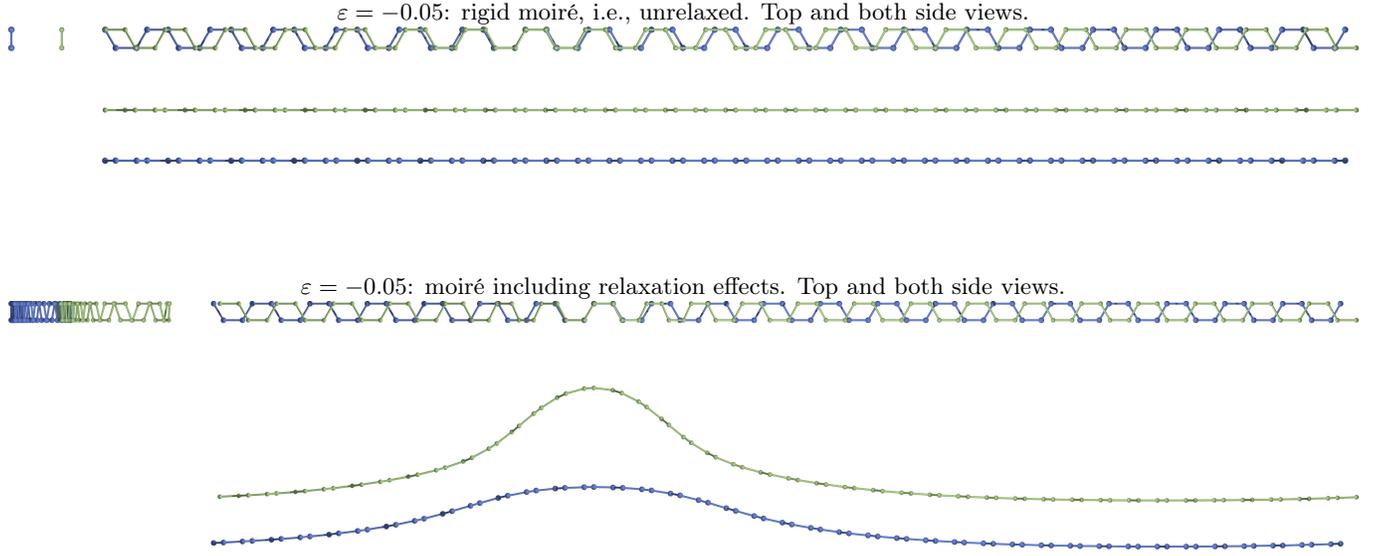

FIG. 12. Structures of rigid and relaxed hBLGs, i.e., without and with relaxation effects.

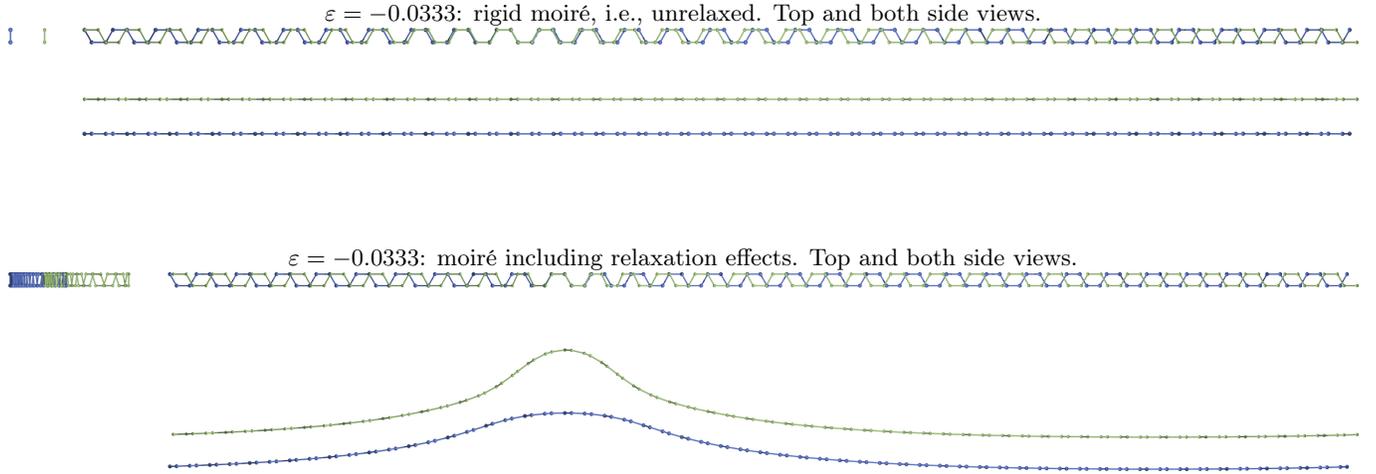

FIG. 13. Structures of rigid and relaxed hBLGs, i.e., without and with relaxation effects.

$\varepsilon = -0.025$: rigid moiré, i.e., unrelaxed. Top and both side views.

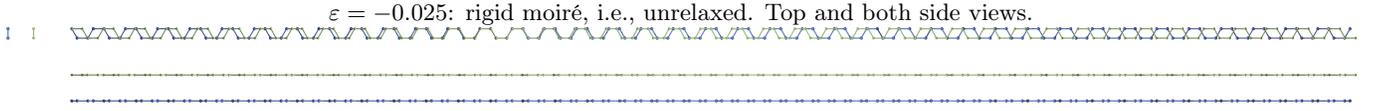

$\varepsilon = -0.025$: moiré including relaxation effects. Top and both side views.

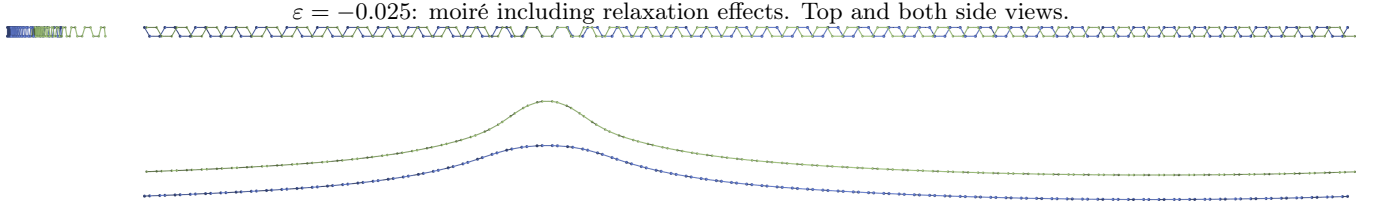

FIG. 14. Structures of rigid and relaxed hBLGs, i.e., without and with relaxation effects.

$\varepsilon = -0.01$: rigid moiré, i.e., unrelaxed. Top and both side views.

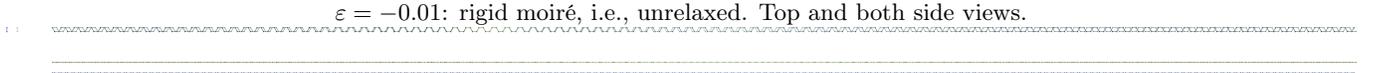

$\varepsilon = -0.01$: moiré including relaxation effects. Top and both side views.

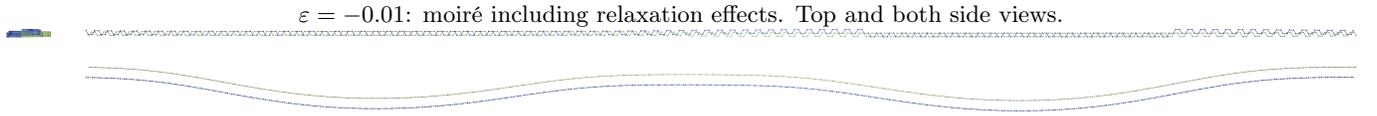

FIG. 15. Structures of rigid and relaxed hBLGs, i.e., without and with relaxation effects.

$\varepsilon = +0.01$: rigid moiré, i.e., unrelaxed. Top and both side views.

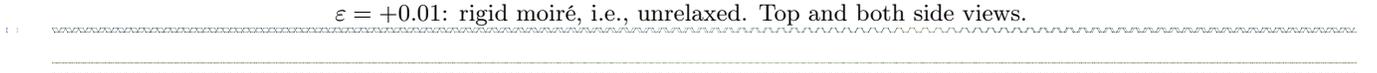

$\varepsilon = +0.01$: moiré including relaxation effects. Top and both side views.

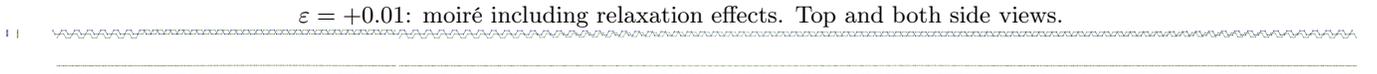

FIG. 16. Structures of rigid and relaxed hBLGs, i.e., without and with relaxation effects.

$\varepsilon = +0.025$: rigid moiré, i.e., unrelaxed. Top and both side views.

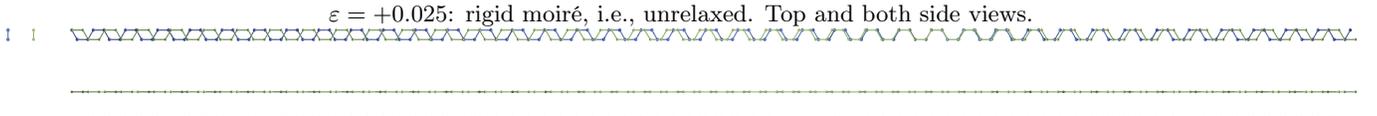

$\varepsilon = +0.025$: moiré including relaxation effects. Top and both side views.

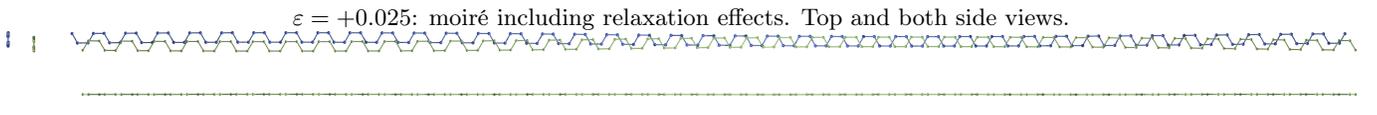

FIG. 17. Structures of rigid and relaxed hBLGs, i.e., without and with relaxation effects.

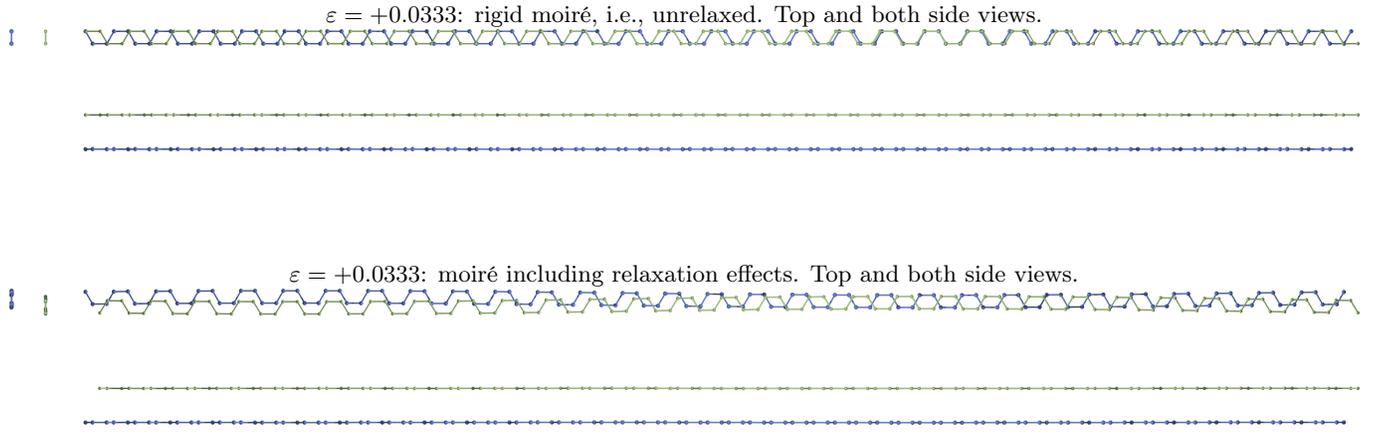

FIG. 18. Structures of rigid and relaxed hBLGs, i.e., without and with relaxation effects.

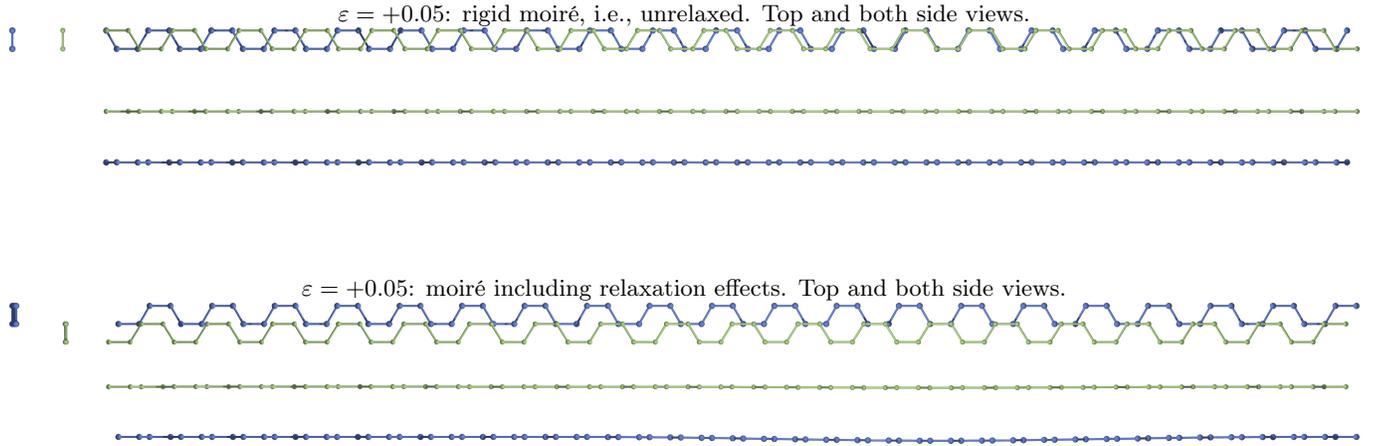

FIG. 19. Structures of rigid and relaxed hBLGs, i.e., without and with relaxation effects.

\* These authors contributed equally

[1] A. P. Thompson, H. M. Aktulga, R. Berger, D. S. Bolintineanu, W. M. Brown, P. S. Crozier, P. J. in 't Veld, A. Kohlmeyer, S. G. Moore, T. D. Nguyen, R. Shan, M. J. Stevens, J. Tranchida, C. Trott, and S. J. Plimpton, LAMMPS - a flexible simulation tool for particle-based materials modeling at the atomic, meso, and continuum scales, Comp. Phys. Comm. **271**, 108171 (2022).

[2] D. W. Brenner, O. A. Shenderova, J. A. Harrison, S. J. Stuart, B. Ni, and S. B. Sinnott, A second-generation reactive empirical bond order (REBO) potential energy expression for hydrocarbons, Journal of Physics: Condensed Matter **14**, 783 (2002).

[3] A. N. Kolmogorov and V. H. Crespi, Registry-dependent interlayer potential for graphitic systems, Physical Review B **71**, 235415 (2005).

[4] F. Gargiulo and O. V. Yazyev, Structural and electronic transformation in low-angle twisted bilayer graphene, 2D Materials **5**, 015019 (2017).

[5] S. Naik, M. H. Naik, I. Maity, and M. Jain, Twister: Construction and structural relaxation of commensurate moiré superlattices 10.48550/arxiv.2102.07884 (2021).

[6] A. García, N. Papior, A. Akhtar, E. Artacho, V. Blum, E. Bosoni, P. Brandimarte, M. Brandbyge, J. I. Cerdá, F. Corsetti, R. Cuadrado, V. Dikan, J. Ferrer, J. Gale, P. García-Fernández, V. M. García-Suárez, S. García, G. Huhs, S. Illera, R. Korytár, P. Koval, I. Lebedeva, L. Lin, P. López-Tarifa, S. G. Mayo, S. Mohr, P. Ordejón, A. Postnikov, Y. Pouillon, M. Pruneda, R. Robles, D. Sánchez-Portal, J. M. Soler, R. Ullah, V. W. zhe Yu, and J. Junquera, Siesta: Recent developments and applications, The Journal of Chemical Physics **152**, 204108 (2020).

[7] J. P. Perdew, K. Burke, and M. Ernzerhof, Generalized gradient approximation made simple, Physical Review B **77**, 3865 (1996).

[8] D. R. Hamann, Optimized norm-conserving vanderbilt pseudopotentials, Physical Review B **88**, 085117 (2013).

[9] A. García, M. J. Verstraete, Y. Pouillon, and J. Junquera, The psml format and library for norm-conserving pseudopotential data curation and interoperability, Computer Physics Communications **227**, 51 (2018).

[10] M. van Setten, M. Giantomassi, E. Bousquet, M. Verstraete, D. Hamann, X. Gonze, and G.-M. Rignanese, The PseudoDojo: Training and grading a 85 element optimized norm-conserving pseudopotential table, Computer Physics Communications **226**, 39 (2018).

[11] J. C. Slater and G. F. Koster, Simplified LCAO method for the periodic potential problem, Physical Review **94**, 1498 (1954).

[12] P. Moon and M. Koshino, Energy spectrum and quantum hall effect in twisted bilayer graphene, Physical Review B **85**, 195458 (2012).

[13] S. Pathak, T. Rakib, R. Hou, A. Nevidomskyy, E. Ertekin, H. T. Johnson, and L. K. Wagner, Accurate tight-binding model for twisted bilayer graphene describes topological flat bands without geometric relaxation, Physical Review B **105**, 115141 (2022).

[14] G. Kresse and J. Hafner, Ab initio molecular dynamics for liquid metals, Physical Review B **47**, 558 (1993).

[15] G. Kresse and J. Furthmüller, Efficient iterative schemes for ab initio total-energy calculations using a plane-wave basis set, Physical Review B **54**, 11169 (1996).

[16] G. Kresse and J. Furthmüller, Efficiency of ab-initio total energy calculations for metals and semiconductors using a plane-wave basis set, Computational Materials Science **6**, 15 (1996).

[17] P. E. Blöchl, Projector augmented-wave method, Physical Review B **50**, 17953 (1994).

[18] M. Bonfanti and R. Martinazzo, Comment on "theoretical study of the dynamics of atomic hydrogen adsorbed on graphene multilayers", Physical Review B **97**, 117401 (2018).


\end{document}